\documentclass[12pt]{iopart}

\usepackage{iopams}
\usepackage{setstack}
\usepackage{graphicx}
\usepackage{amsthm}

\newcommand{\be}{\begin{equation}}
\newcommand{\ee}{\end{equation}}
\newcommand{\bea}{\begin{eqnarray}}
\newcommand{\eea}{\end{eqnarray}}
\newcommand{\nn}{\nonumber}

\begin{document}

\title[An Ocean Drum: quasi-geostrophic energetics from a Riemann geometry perspective]{
An Ocean Drum: 
quasi-geostrophic energetics from a Riemann geometry perspective}

\author{Jos\'e Luis Jaramillo}

\address{
Institut de Math\'ematiques de Bourgogne (IMB), \\
UMR 5584, CNRS, Universit\'e de Bourgogne Franche-Comt\'e, F-21000 Dijon, France}
\address{
Laboratoire de Physique des Oc\'eans (LPO), \\
UMR 6523, CNRS, Universit\'e de Bretagne Occidentale, 29200 Brest, France}
\ead{Jose-Luis.Jaramillo@u-bourgogne.fr}

\begin{abstract}
We revisit the discussion of the energetics of quasi-geostrophic flows 
from a geometric perspective based on the introduction of an 
effective metric,  built in terms of the flow stratification and 
the Coriolis parameter. In particular,  an appropriate notion of normal modes is
defined through a spectral geometry problem in the ocean basin (a compact 
manifold with boundary) for the associated Laplace-Beltrami scalar  operator. This spectral 
problem can be used to systematically encode non-local aspects of
stratification and topography. 
As examples of applications we revisit the isotropy assumption
in geostrophic turbulence,  identify (a patch of) the hyperbolic 
space $\mathbb{H}^3$ as the leading-order
term in the effective geometry for the deep mesoscale ocean and,
finally, discuss some diagnostic tools  based on a simple
statistical mechanics toy-model to be used in numerical simulations and/or 
observations of quasi-geostrophic flows.

\end{abstract}

%
\vspace{2pc}
\noindent{\it Keywords}: quasi-geostrophic equations,  
spectral geometry, statistical mechanics
%
%
%
%

\section{Introduction}
\label{s:Introduction}
Quasi-geostrophic (QG) dynamics describes some of the slow motions of the ocean and atmosphere
at mid-high latitudes. Our main goal in this article is to 
recast QG dynamics in a form appropriate for the application 
of general Riemann geometrical analysis tools, that
can bring insight into generic qualitative aspects of the QG model.

Our main ultimate motivation is the assessment of
generic features of stratification and topography in ocean dynamics, 
in particular in QG dynamics in the mesoscale ocean  ($\sim 20-200$ km) \cite{Gill82,pedlosky:1990,Salmon98,Vallis06,cushman2011}. 
QG dynamics applies generically to rotating (stratified) fluids
close to geostrophic equilibrium (characterized by the balance between Coriolis accelerations
and horizontal pressure gradients), where horizontal scales are much larger than
vertical ones. QG dynamics can then be seen as a
quasi-static transition along instantaneous geostrophic equilibria, that filters
rapid motions such as internal waves while keeping sufficient dynamical richness
to study the slow motions. The latter domi-nate the energy content
at low (subinertial) frequencies \cite{FerWun09}.
The QG model presents remarkable structural properties, 
namely the conservation of physical quantities such as the energy and enstrophy, that permit
to account for non-trivial observed physical phenomena and work as a test-bed for
more complex treatments of ocean dynamics.
Of particular relevance is the discussion of 
geostrophic turbulence \cite{Charney71,Rhines79,Salmon80}
and its role in the formation of large coherent structures (geostrophic eddies, jets, filaments)
through the associated turbulent cascades, similar to the 
$2$-dimensional turbulence mechanism \cite{Kraich67,Leith68,Batche69}.
This explains the success of the QG model, 
even if it offers a very simplified treatment of ocean dynamics. In spite of the extensive study
of QG dynamics, we argue here that there is still room for further refinement 
by incorporating generic qualitative aspects encoded in 
the geometric structure of the operators controlling the dynamics.

In the present article we adopt such a geometric approach and start
exploring the insights it provides into some qualitative features
of ocean mesoscale dynamics.
In a first stage, we introduce the basic elements 
to revisit  QG dynamics from a 
Riemann geometry perspective.
In particular, we construct an effective (Riemann)
metric from the stratification
and Coriolis ingredients of a given (bulk) water mass and
recast QG equations in terms 
of the associated (exact) $3$-dimensional scalar Laplacian. 
We focus then on QG energetics and formulate a 
Laplacian eigenvalue problem in the ocean basin, where
topography enters through the boundary conditions.  
This spectral geometry problem, that incorporates explicitly
stratification and topography elements of the basin (and that we refer to as an 
{\em ocean drum})  opens an avenue to the qualitative analysis 
of the QG model: spectral geometry tools permit to extract
information of the system without the need of explicitly solving 
the corresponding analytic equations.
The presentation of these formal geometric elements constitutes the main goal of this
article.

In a second stage, at a more exploratory level, we consider 
some lines of application in the ocean context (the formal elements above
essentially apply also in an atmosphere setting). First, we discuss
how the
geometric perspective brings further insight into the 
isotropy assumption in the analysis of geostrophic turbulence, in particular
addressing the optimal stratification conditions for its occurrence. 
Second, we point out the potential role of hyperbolic geometry elements
in the study of QG structures in the deep ocean, opening an ocean physics scenario
for the transfer and application of geometric results in Riemann spaces of 
(constant) negative curvature. 
Third,  we present a set of diagnosis 
tools for the analysis of observational data/numerical simulations
of QG flows. Concretely, these qualitative estimators are constructed
out of an (ad hoc) statistical mechanics toy model motivated 
by the ocean drum problem. The specific goal of such a model is not to capture the 
actual QG physics, but rather to provide a  simple approach
to map QG field configurations, namely taking into account stratification and topography, 
through the systematics of the related thermodynamical quantities. The previous three examples 
constitute a brainstorm exercise on applications of the 
adopted geometric approach. Regarding a more systematic framework, 
the Robert-Sommeria-Miller (RSM) \cite{Robert90,Miller90,Robert91,RobSom91} 
equilibrium statistical mechanics theory for $2-$dimensional flows provides
a powerful formalism for the application of the (spectral) Riemann geometry elements
presented here, in particular in the construction of invariant measures for 
continuously stratified QG flows along the lines in \cite{BouCor10}.
This statistical mechanics objective is an ultimate target of our approach, requiring a specific discussion 
to be addressed elsewhere.

The plan of the article is the following. Section
\ref{s:QGdynamics_energetics} briefly reviews the fundamentals
of QG dynamics and energetics. Section \ref{s:RiemannGeometry}
presents the elements of Riemann geometry needed
in the rest of the article.  These first two sections 
introduce the basic building blocks, in particular setting the notation. 
Section \ref{s:QGRiemann} contains 
the main contribution of the article, namely the identification
of an effective metric permitting a compact geometric rewriting 
of the QG elements, in particular through the formulation of 
a geometric eigenvalue problem (ocean drum) for a given ocean basin. 
In section 
\ref{s:OceanPhysics}, and in contrast with the 
``mathematical physics'' flavour of section \ref{s:QGRiemann}, we adopt a more heuristic approach
and apply (in a ``theoretical physics'' spirit) the previous
geometric elements to the specific discussion in the ocean.
Conclusions are presented in section \ref{s:Conclusions}.
The main text is complemented with five technical appendices. 
Having in mind a broad reader profile,
we make an effort to present an essentially 
self-contained discussion, perhaps
at the price of revisiting elements occasionally too basic for
the expert in the field.

\section {Quasi-geostrophic dynamics and energetics}
\label{s:QGdynamics_energetics}
Let us consider an ocean basin manifold $M$ with given Cartesian
coordinates $(x, y, z)$ and an incompressible flow 
$\mathbf{v}$, with components 
$\mathbf{v}= u \,\mathbf{e}_x + v \,\mathbf{e}_y + w\,\mathbf{e}_z$ in the 
Cartesian orthonormal frame. 
The QG equations for a rotating and stratified fluid
at mid and high latitudes are derived from the Navier-Stokes equation in Boussinesq approximation,
in an (asymptotic) expansion around 
the hydrostatic and geostrophic equilibria. 
As stated above, geostrophic equilibrium is achieved by the 
balance between rotation (Coriolis) accelerations and horizontal pressure gradients.
The geostrophic flow is
characterized by a velocity $\mathbf{v}_g$, expressed (in the so-called
f-plane approximation)  as (see e.g. \cite{pedlosky:1990})
\bea
\label{e:geostrophic_velocity}
\mathbf{v}_g = \mathbf{e}_z\times \nabla \psi \ , 
\eea
where the streamfunction $\psi$ is given by 
$\psi=\frac{p}{f_o\rho_o}$, with $p$ the pressure (anomaly), $\rho_o$
a mean reference value for the density and $f_o$
the Coriolis parameter at latitude $\varphi_o$
(more generally, $f=2\Omega \sin\varphi$, with
$\Omega$ the Earth's rotation frequency and $\varphi$ a given
latitude). Therefore the geostrophic $\mathbf{v}_g$ defines,
at each depth $z$, a $2$-dimensional divergence-free flow.
QG dynamics is obtained as a small deviation 
of the geostrophic flow in a regime of
fast environmental rotation and strong density stratification. These features
are respectively characterized by the smallness of the Rossby
($\mathrm{Ro}$) and Froude ($\mathrm{Fr}$) numbers. 
The former is defined in terms of a characteristic horizontal velocity
$U$, the rotation frequency $f_o$ and a horizontal length scale $L$ 
as $\mathrm{Ro}=\frac{U}{f_o L}$, whereas the latter 
is given by $\mathrm{Fr}=\frac{U}{N H}$, where $H$ is a vertical length scale and
$N$ is the buoyancy or Brunt-V\"ais\"al\"a frequency 
accounting for the density stratification ($N^2=-\frac{g}{\tilde{\rho}}\frac{d\tilde{\rho}}{dz}$,
with $g$ the gravitational acceleration and $\tilde{\rho}(z)$ an average
density at depth $z$, providing the reference stratification around which the 
density $\rho$ is linearized: $\rho=\tilde{\rho}+ \delta\rho$ with $\delta\rho/\tilde{\rho}\ll 1$).
Under the assumption $\mathrm{Ro} \sim \mathrm{Fr} = \epsilon\ll 1$ 
and expressing the velocity as $\mathbf{v}=\mathbf{v}_g + \mathbf{v}_{ag}$, 
where $\mathbf{v}_{ag}\sim o(\epsilon)$ denotes the corrections
to the geostrophic velocity, QG equations are
obtained from an expansion in $\epsilon$ of an appropriate
rescaling of the rotating 
Boussinesq equations (for formal details of the derivation and further physical insight, 
including the whole set of assumptions in the QG model,
see e.g. \cite{Gill82,pedlosky:1990,Salmon98,Vallis06,majda03,cushman2011}). The resulting QG
dynamics filters the fast components of the flow, describing the slow
motions captured by the conservation
of a scalar quantity $q$ referred to as the {\em QG potential vorticity}
\bea
\label{e:potential_vorticity}
q = \Delta_\parallel\psi + \frac{1}{\tilde{\rho}}\partial_z\left(\tilde{\rho}\frac{f^2_o}{N^2}
\partial_z\psi\right) 
+ f \ , 
\eea
with $\Delta_\parallel = \frac{\partial^2}{\partial x^2} + \frac{\partial^2}{\partial y^2}$.
In the absence of forcing and dissipation (that can be neglected 
without loss of generality in the present discussion) QG dynamics is specified by
\bea
\label{e:QG_dynamics}
\partial_t q + v_g^i \nabla_i q = 0 \ ,
\eea
with $v_g^i$ the components of the geostrophic velocity in (\ref{e:geostrophic_velocity}).
It is customary to write it as
\bea
\label{e:QG_dynamics_2}
\partial_t q + J(\psi,q) = 0 \ ,
\eea
where, $J(\psi,q)=\partial_x\psi\partial_yq - \partial_y\psi\partial_xq$. 
The state of the flow is therefore encoded in the 
streamfunction $\psi$, whose dynamics is dictated by a geostrophic 
$2$-dimensional advection balancing 
the (vertical component of the) relative vorticity
($\zeta=\partial_xv-\partial_y u= \Delta_\parallel\psi$), 
a vertical-stretching contribution
and the planetary vorticity term $f$
[respectively the first, second and third terms in the  right-hand-side
in Eq. (\ref{e:potential_vorticity})]. QG equations
are completed with appropriate boundary conditions, in particular a
vanishing vertical (ageostrophic) velocity\footnote{The first contribution to the (ageostrophic)
vertical velocity is $w = \frac{f_o}{N^2}\left[\frac{\partial^2\psi}{\partial t \partial z}
+ J(\psi, \partial_z\psi) \right]$ (e.g. \cite{cushman2011}).}, leading to 
the geostrophic advection of $\partial_z\psi$ on the boundary ${\partial M}$ (e.g. \cite{cushman2011})
\bea
\label{e:vertical_BC}
\left[\partial_t \left(\partial_z\psi\right) + J(\psi,\partial_z\psi) \right]|_{\partial M}= 0 \ .
\eea
This condition is actually devised for the ocean bottom and surface. When considering
restricted ocean domains, lateral boundary conditions are also needed. A common  choice
involves a Dirichlet condition only depending on time, $\psi|_{\partial M} = C(t)$. Topography
effects are a main interest in the present work, this leading to a focus 
on condition (\ref{e:vertical_BC}) (cf. however the remark on lateral boundary conditions  
after the eigenvalue problem in \ref{s:AnOceanDrum}).

Energetics are of particular relevance in our discussion.
The energy $E_{QG}[\psi]$ of a QG flow with streamfunction $\psi$
contains two terms (e.g. \cite{pedlosky:1990}): 
the kinetic energy of the flow,
$E_{\mathrm{kin}}[\psi]= \frac{1}{2}\int_M |\mathbf{v}_g|^2 \tilde{\rho}d^3x
=\frac{1}{2}\int_M |\nabla_\parallel \psi|^2 \tilde{\rho}d^3x$, 
and a second term accounting for the so-called
{\em available potential} (AP)  energy, of the form $E_{\mathrm{AP}}[\psi]=\frac{1}{2}\int_M 
\frac{f^2_o}{N^2} (\partial_z \psi)^2  \tilde{\rho} d^3x$. 
They correspond  respectively to the relative vorticity and vertical-stretching 
terms
\bea
\label{e:QGenergy_1}
E_{QG}[\psi] =   E_{\mathrm{kin}}[\psi] + E_{\mathrm{AP}}[\psi]
= \frac{1}{2}\int_M \left(|\nabla_\parallel \psi|^2
+ \frac{f^2_o}{N^2} (\partial_z \psi)^2  \right)  \tilde{\rho} d^3x \ .
\eea
Notably the planetary vorticity plays no role in the energetics.
Indeed, only the $q-f$ part of the QG
potential vorticity $q$ enters into the energy, defining
a linear elliptic operator $\hat{L}$ acting on $\psi$, 
introduced by Charney \cite{Charney71} 
in the analysis of geostrophic turbulence 
\bea
\label{e:Charney_operator}
\hat{L} \psi \equiv  q-f =\Delta_\parallel \psi + \frac{1}{\tilde{\rho}}\partial_z\left(\tilde{\rho}
\frac{f^2_o}{N^2}\partial_z\psi \right) \ .
\eea
In terms
of the operator $\hat{L}$ the QG energy can be rewritten
(up to boundary terms) as
\bea
\label{e:QGenergy_2}
\!\!\!\!\!\!E^{\mathrm{T}}_{QG}[\psi] = 
- \frac{1}{2}\int_M  \psi 
\left[\Delta_\parallel + \frac{1}{\tilde{\rho}}\partial_z\left(\tilde{\rho}\frac{f^2_o}{N^2}
\partial_z\right)\right]\psi\; \tilde{\rho}d^3x  
= - \frac{1}{2}\int_M  \psi \hat{L} \psi \; \tilde{\rho} d^3x \ .
\eea
This form of the energy as the expectation value of the operator $\hat{L}$
at the ``state'' given by the streamfunction $\psi$ presents 
two appealing features.
On the one hand, from a technical perspective it suggests
to explore the energetics associated with $E_{QG}[\psi] $  in terms of 
the analysis of the operator $\hat{L}$, in particular through
the systematic study of its spectral properties
(the analysis of cascades in geostrophic turbulence \cite{Charney71} 
is akin to such approach).
On the other hand, from a heuristic
point of view, the formal resemblance with the energy expectation values
in quantum mechanics can be exploited to introduce novel approaches
into the problem, complementary to more standard fluid mechanics treatments.
As an example, the calculation of energy corrections to a dynamical QG flow, from a given reference one,
can benefit from tools in operator perturbation theory.

\subsection{Enstrophy}
Let us define the enstrophy ${\cal E}[\psi]$ of a QG flow, given by the streamfunction
$\psi$, as
\bea
\label{e:enstrophy_QG}
{\cal E}[\psi] = \int_M \left(\hat{L}\psi\right)^2 \tilde{\rho}\,d^3x \ .
\eea
A key feature of QG dynamics is the conservation of ${\cal E}[\psi]$
along the QG flow. The existence of this additional invariant plays 
a key role in geostrophic
turbulence \cite{Charney71}. More generally, functionals
of the potential vorticity $q$ associated with arbitrary functions $s$
\bea
\label{e:Casimir}
{\cal C}_s[q] = \int_M s(q) d^3x \ ,
\eea
are also preserved by the QG flow. These infinite dynamical invariants, referred to as Casimirs,
play a fundamental role in the statistical mechanics of QG flows (e.g. \cite{BouchetVenaille12}).

\section{Riemannian geometry and Laplace-Beltrami operator}
\label{s:RiemannGeometry}
In this section we collect the elements of Riemannian geometry
that we need for our discussion, in particular fixing the notation
(see additional details in \ref{s:Appendix_coordinateexpressions}).

\subsection{Some elements of Riemannian geometry}
\subsubsection{Metric tensor.}
Let us consider a Riemannian metric $\mathbf{g}$ defined on a compact
$d$-dimensional manifold $M$, possibly with boundaries
(we shall focus on $d=3$, but we keep a general $d$ in this section). 
Given a local chart $\{x^{i_1},\dots,x^{i_d}\}$,  the metric
is written as $\mathbf{g}=g_{ij}dx^i\otimes dx^j$, where $g_{ij}$ is a 
non-degenerate positive-definite symmetric matrix. 
The metric line element is written as $ds^2= g_{ij}dx^idx^j$,
with $s$ the arc-length parameter.
Let us denote the inverse of the $g_{ij}$ matrix by $g^{ij}$, 
i.e. $g^{ik}g_{kj}={\delta^i}_j$, and its determinant by 
$g=\det(g_{ij})$. The metric volume form is  
given by $\boldsymbol{\epsilon}_g=\sqrt{g}\,dx^1\wedge\dots\wedge dx^d
=\frac{\sqrt{g}}{d!}\,\epsilon_{i_1\dots i_d } dx^{i_1}
\wedge\dots\wedge dx^{i_d}$,
with $\epsilon_{i_1\dots i_d}$ the completely antisymmetric tensor with $\epsilon_{1...d}=1$.

\subsubsection{Connection and curvature.}
\label{s:connectioncurvature}
Let us denote by $\boldsymbol{\nabla}$ the Levi-Civita connection associated
with $\mathbf{g}$, i.e. the unique connection that is 
metric compatible (i.e. $\nabla_k g_{ij}=0$) and torsion free 
(i.e. $\left(\nabla_i\nabla_j- \nabla_j\nabla_i\right)\phi = 0$, for all scalar 
fields $\phi$ on $M$). Geodesics satisfy $u^j\nabla_j u^i=0$, with 
unit-tangent vector $\mathbf{u}=u^i\partial_{x^i}=\frac{dx^i}{ds}\partial_{x^i}$.
Regarding its curvature, the Riemann tensor components are denoted by
${R^i}_{jkl}$, where $\left(\nabla_k\nabla_l - \nabla_l\nabla_k\right)V^i = {R^i}_{jkl}V^j$,
with $\mathbf{V}=V^i\partial_{x^i}$ any vector field on $M$. The 
Ricci tensor is the trace of the Riemann tensor, with components
$R_{ij}={R^k}_{ikj}$, and 
the Ricci scalar is $R={R^i}_i=g^{ij}R_{ij}$.

\subsubsection{Geometry of the boundary $\partial M$.} 
When considering topography, it becomes relevant to 
control the geometry of the boundary of $M$.
Let us consider the boundary $\partial M$ with outgoing
normal vector $\mathbf{s}=s^i\partial_{x^i}$, 
normalized as $s^is_i=g_{ij}s^is^j=1$.
Let us denote by $\mathbf{h}$ the Riemannian metric induced on  $\partial M$
from $\mathbf{g}$. We denote also by $\mathbf{h}$
the orthogonal projector onto $\partial M$, 
given by $\mathbf{h}={h^i}_j \partial_{x^i}\otimes dx^j
=\left({g^i}_j - s^i s_j\right)\partial_{x^i}\otimes dx^j 
= \left({\delta^i}_j - s^i s_j\right)\partial_{x^i}\otimes dx^j$.
Denoting by $h$ the determinant of $\mathbf{h}$, 
the volume (area) form on $\partial M$ is 
$\boldsymbol{\epsilon}_h=\sqrt{h}\,dx^1\wedge\dots\wedge dx^{d-1}$
(this assumes $\partial M$ given by $x^d=\mathrm{const}$).
The Levi-Civita connection of $\mathbf{h}$ is
denoted by $\mathbf{D}$, i.e. $D_kh_{ij}=0$ and 
$\left(D_iD_j- D_jD_i\right)\phi = 0$.
The extrinsic curvature of $(\partial M, \boldsymbol{h})$ into 
$(M, \boldsymbol{g})$ is given by the tensor
$\mathbf{K} = \left({h^k}_i{h^l}_j\nabla_k s_l\right) dx^i\otimes dx^j$,
defined on $\partial M$. The mean curvature $K$ of $\partial M$
is the trace of $\mathbf{K}$, i.e. $K = h^{ij}K_{ij}=\nabla_i s^i$.

\subsubsection{Laplacian operator.}
A fundamental object in our discussion is the scalar Laplace-Beltrami operator
$\Delta_g$ associated with the metric $\mathbf{g}$, that we shall refer to
simply as the Laplacian.
The action of $\Delta_g$ on a scalar $\phi$, 
is given by the divergence of its gradient
\bea
\label{e:Laplacian_g}
\Delta_g \phi = \nabla_i \nabla^i \phi = g^{ij} \nabla_i\nabla_j \phi \ .
\eea
 Given arbitrary coordinates, it holds the important expression
in our present context 
\bea
\label{e:Laplacian_g_partial}
\Delta_g \phi = \frac{1}{\sqrt{g}}\partial_{x^i}\left(\sqrt{g}\;g^{ij}\partial_{x^j} 
\phi  \right) \ .
\eea

\subsubsection{Integration on $M$ and $\partial M$.} 
The volume form $\boldsymbol{\epsilon}_g$ permits
to integrate a scalar $\phi$ on $M$, as $\int_M \phi\, \boldsymbol{\epsilon}_g$.
As notation, we introduce the associated integration measure as $\mu_g = dV_g = \sqrt{g} \,d^dx$.
In particular the volume $V_g$ of $M$, as measured by $\mathbf{g}$, is
\bea
\label{e:volume_M}
V_g = \int_M 1 \cdot \boldsymbol{\epsilon}_g
=  \int_M \sqrt{g} \,d^dx = \int_M dV_g\ .
\eea
A scalar product $\langle \cdot, \cdot\rangle_g$ is defined on $L^2(M, \mu_g)$, as
\bea
\label{e:scalar_product}
\langle \phi | \psi \rangle_g \equiv  \int_M \phi^* \psi \; dV_g
=\int_M \phi^* \psi  \;\sqrt{g} d^dx \ ,
\eea
for generally complex scalar functions $\phi$ and $\psi$.
In particular, the Laplacian $\Delta_g$ is self-adjoint when restricted
to functions satisfying homogeneous Dirichlet or Neumann conditions (more generally,
homogeneous Robin conditions). Given a scalar $\phi$ we define the
quantity $E_g[\phi]$, referred to as the {\em energy} of $\phi$ (also
{\em Dirichlet energy} \cite{Chavel84}), as 
\bea
E_g[\phi] \equiv \int_M \nabla_i \phi^*  \nabla^i \phi \,dV_g = \int_M g^{ij}\nabla_i \phi^*  \nabla_j \phi \,dV_g 
= \int_M |\mathbf{\nabla} \phi|_g^2 \,dV_g  .
\eea
We denote by $E^{\mathrm{T}}_g$ the expected value
of the operator $(-\Delta_g)$ at the ``state'' $\phi$, i.e. 
\bea
E^{\mathrm{T}}_g[\phi]  \equiv \langle \phi | (-\Delta_g)| \phi \rangle_g =
\int_M \phi^* \left(-\Delta_g\right) \phi \, dV_g \ .
\eea
As for $M$, we integrate on $\partial M$ with the volume
form $\boldsymbol{\epsilon}_h$. Specifically, we introduce 
the integration measure $dA_h=\sqrt{h}\,dx^{d-1}$. Then we define the
quantity $E_{\partial M}[\phi]$ as
\bea
E^{\partial M}_g[\phi] \equiv - \int_{\partial M} \phi^* s^i\nabla_i \phi \, dA_h  \ ,
\eea
that we shall refer to as {\em boundary energy} of $\phi$.
Integrating by parts it follows the relation 
\bea
\label{e:relation_Energies}
E^{\mathrm{T}}_g[\phi] = E_g[\phi] + E^{\partial M}_g[\phi] \ .
\eea
One could refer to  $E_g[\phi]$ as the {\em bulk energy} of $\phi$
and then to $E^{\mathrm{T}}_g[\phi]$ as its {\em total energy}.

\subsection{The spectrum of the Laplacian}
Spectral geometry offers a framework for studying
geometrical properties of a manifold $M$ out of 
the spectrum of specific differential
operators defined on $M$ and, the other way around, to address
generic properties of operators' spectra 
from the a priori geometric knowledge of $M$.
The spectral study of the 
Laplacian on compact Riemannian
manifolds has been systematically addressed,  the
drum's problem \cite{Kac66,GirTha10} being a prototype.

We consider the eigenvalue
problem of the Laplacian on a compact Riemannian 
manifold with boundaries, subject to homogeneous Neumann 
boundary conditions
\bea
\label{e:spectral_Delta}
-\Delta_g \phi_n = \lambda_n \phi_n  \ \ \ , \ \ \ 
\left.\partial_{\mathbf{s}}\phi_n\right|_{\partial M}\equiv 
\left.s^i\nabla_i \phi_n \right|_{\partial M} = 0 \ .
\eea
More generally, we could consider a mixed eigenvalue problem  with
(\ref{e:spectral_Delta}) in part of $\partial M$ and 
homogeneous Dirichlet conditions in the rest of $\partial M$ \cite{Chavel84}.
We will focus here on the strict Neumann case (cf. \ref{s:AnOceanDrum}).
Eigenvalues $\lambda_n$ are non-negative and can be ordered as 
\bea
\label{e:lambda_ordered}
0=\lambda_0 <\lambda_1 \leq \lambda_2 \leq \dots \leq \lambda_n \leq \dots \to \infty\ ,
\eea 
with the associated $\{\phi_n\}$ a complete set of orthogonal eigenfunctions on $L^2(M, \mu_g)$
\cite{Chavel84}.

The exact eigenvalue problem is generically out of reach.
The power of the geometric approach is that generic qualitative 
results can be established, even if an explicit solution is absent.
We collect some results of interest (see \cite{Chavel84}, and 
\cite{Berger03} for a pedagogical review):
\begin{itemize}
\item[i)] {\em Bounds on low eigenvalues}: lowest non-vanishing eigenvalues encode 
large scale information. In particular, lower and upper bounds for  
$\lambda_1$ are given by 
\bea
\label{e:bounds_lower_eigenvalues}
\lambda_1 \geq \frac{1}{4}h_c^2 \ \ , \ \  \lambda_1 \lesssim 1/V_g^{\frac{2}{d}} \ ,
\eea
where $h_c$ is the so-called Cheeger constant~\cite{Chavel84}, essentially controlling the largest
``diameter'' of $M$. A useful variational characterization of $\lambda_1$
is given by the Rayleigh quotient [with $\phi\in C^\infty(M)$, or more
generally the Sobolev space $H^1(M)$]
\bea
\label{e:Rayleigh_quotient}
\lambda_1 = 
\inf_{(\phi\neq 0, \int_M \phi =0)} 
\frac{\int_M \nabla^i \phi \nabla_i \phi \; dV_g}{\langle \phi|\phi \rangle_g} 
= \inf_{(\phi\neq 0, \int_M \phi =0)}
\frac{E_g (\phi)}{\langle \phi|\phi \rangle_g} \ .
\eea
The minimum is reached for $\phi=\phi_1$ and the characterization
can be extended to higher eigenvalues $\lambda_n$ by imposing orthogonality of
$\phi$ and $\phi_{i<n}$.

\item[ii)] {\em Weyl's law}: regarding high-eigenvalue asymptotics and
denoting with $N(\lambda)$ the number of eigenvalues
$\lambda_n$ (including multiplicity) such that $\lambda_n\leq \lambda$,
it holds
\bea
\label{e:Weyl_law}
N(\lambda) \sim  \frac{B_d}{(2\pi)^d}  V_g \lambda^{\frac{d}{2}} \ \ (\lambda\to\infty) \ ,
\eea
with $B_d$ the volume of the unit ball in $\mathbb{R}^d$. The
asymptotic expression follows
\bea
\label{e:asymtoptics_lambda_k}
\lambda_n \sim \frac{(2\pi)^2}{B_d^{\frac{2}{d}}} \left(\frac{n}{V_g}\right)^{\frac{2}{d}} 
\ \ (n\to\infty)\ .
\eea
\item[iii)] {\em Heat kernel expansion}: spectral functions of an
operator, defined from its spectrum 
and a (generically complex) parameter, offer a powerful way
of encoding operator information.
For the Laplacian, a prominent example is given by the 
{\em heat kernel}
\bea
K(t)=\sum_n^\infty e^{-t\lambda_n} \ \ ,
\eea
namely the trace of the heat equation Green's function.
Its asymptotics for $t\to 0$ can be obtained
without an explicit knowledge of the spectrum. It holds \cite{BransonGilkey90,Gilkey2003}
\bea
\label{e:Heat_Kernel}
\!\!\!\!\!\!\!\!\!\!\!\!\!\!\!K(t) =  \frac{1}{(4\pi t)^{\frac{d}{2}}}
\left(V_g +  \frac{\sqrt{\pi}}{2} A_h t^\frac{1}{2} + a_1 t + b_1 t^\frac{3}{2} 
+ a_2 t^2 + \dots \right) \ \ \ (t\to 0) \ .
\eea
The heat kernel coefficients $a_i$ and $b_i$ are completely characterised
by the geometry of $M$ and $\partial M$, depending on boundary conditions
in the spectral problem (\ref{e:spectral_Delta}) [see \ref{s:HK_coefficients}
for expressions of the first terms  \cite{BransonGilkey90,Gilkey2003},
in the Neumann case in (\ref{e:spectral_Delta})].

\end{itemize}

\section{Quasi-geostrophy from a Riemannian perspective}
\label{s:QGRiemann}
Charney's operator $\hat{L}$ in (\ref{e:Charney_operator}) is essentially a Laplacian
(indeed, an exact ``flat'' Laplacian for constant stratification). This plays an 
important role in geostrophic turbulence \cite{Charney71,Vallis06}.  
In the light of the above-presented geometric Riemannian elements, 
we show that this is also the case for an arbitrary stratification
(with $z$, but also $(x,y)$ dependence). 
This leads us to revisit some aspects of the QG 
dynamics from a Riemannian perspective.

\subsection{Quasi-geostrophic metric}
Let us introduce an effective Riemannian metric $\mathbf{g}$
in an ocean basin $M$, with the aim of relating the operator $\hat{L}$ 
in (\ref{e:Charney_operator}) to the corresponding
metric Laplacian  $\Delta_g$. Considering standard coordinates $(x, y, z)$, we 
start by adopting the following Ansatz for the metric 
\bea
\label{e:metric_Ansatz}
ds^2 = a^2\left(dx^2 + dy^2\right) + b^2 dz^2 \ \ , \ \  \hbox{that is} \ \ 
g_{ij} =\left(\begin{array}{ccc} a^2 & 0 & 0 \\ 0 & a^2 & 0 \\ 0 & 0 & b^2 \end{array}\right) \ ,
\eea
where $a$ and $b$ depend on  $(x,y,z)$. The inverse matrix 
$g^{ij}$ and the determinant are
\bea
g^{ij} =\left(\begin{array}{ccc} a^{-2} & 0 & 0 \\ 0 & a^{-2} & 0 \\ 0 & 0 & b^{-2} \end{array}\right)
\ \ \ , \ \ \ g = a^4b^2 \ \ \ , \ \ \ \sqrt{g} = a^2b \ .
\eea
Choosing for $a$ and $b$ the expressions
\bea
\label{e:a_b}
a = \left(\frac{\tilde{\rho}}{\rho_o}\right) \left(\frac{f_o}{N}\right) 
\ \ , \ \ b = \frac{\tilde{\rho}}{\rho_o} \ ,
\eea
the determinant becomes 
\bea
\label{e:determinant_g}
\sqrt{g} =  \left(\frac{\tilde{\rho}^3}{\rho_o^3}\right)\left(\frac{f_o^2}{N^2}\right) \ \ ,
\eea
so that, inserting these elements in the expression (\ref{e:Laplacian_g}) for $\Delta_g$,
we can rewrite $\hat{L}$ as
\bea
\label{e:L_Delta_g_det}
\hat{L} \psi = \left(\frac{\tilde{\rho}^2}{\rho_o^2}\right)\left(\frac{f_o^2}{N^2}\right)
\Delta_g\psi \  =  \left(\frac{\rho_o}{\tilde{\rho}}\right)
\Delta_g\psi  \; \sqrt{g} \ .
\eea 
In the generic case, there is no choice of functions $a$ and $b$ making 
$\hat{L}$ the Laplacian of some $\mathbf{g}$ (again, for constant 
$\tilde{\rho}$ and $N$ this is the flat Euclidean
Laplacian, with a constant coordinate rescaling). However, the factor is precisely
the good one for the energetics. Indeed, inserting (\ref{e:L_Delta_g_det})
in  (\ref{e:QGenergy_2}), the energy $E^{\mathrm{T}}_{QG}$
is exactly recast as an expectation value (for the streamfunction $\psi$)
of the Laplace operator $\left(-\Delta_g\right)$
associated with $\mathbf{g}$, with the natural Riemannian integration measure $dV_g=\sqrt{g}d^3x$.
This proves the following Lemma.

\medskip

\noindent{\bf Lemma 1 (quasi-geostrophic metric)}. {\em Given a QG flow
in a domain $M$ with Cartesian coordinates $(x,y,z)$, 
streamfunction $\psi$, 
density stratification $\tilde{\rho}$ with  Brunt-V\"ais\"al\"a frequency $N$ and Coriolis parameter 
$f_o$, let us define the {\em effective QG metric}} as
\bea
\label{e:QG_effective_metric}
ds^2 = \left(\frac{\tilde{\rho}^2}{\rho_o^2}\right)
\left(\frac{f_o^2}{N^2} \left(dx^2 + dy^2\right) + dz^2\right) \ .
\eea
{\em Then Charney's operator $\hat{L}$ is written as}
\bea
\label{e:L_Delta_g}
\hat{L} \psi = \left(\frac{\tilde{\rho}^2}{\rho_o^2}\right)\left(\frac{f_o^2}{N^2}\right)
\Delta_g\psi \ ,
\eea 
{\em and the QG energies $E_{QG}$ and $E^{\mathrm{T}}_{QG}$ for the
streamfunction $\psi$ are respectively given by}
\bea
\label{e:energy_gradient}
E_{QG}[\psi] 
= \frac{1}{2}\rho_o\int_M \nabla^i\psi \nabla_i\psi \, dV_g 
= \frac{1}{2}\rho_o\int_M |\mathbf{\nabla}\psi|_g^2 \, dV_g \ , 
\eea
{\em and}
\bea
\label{e:total_energy_Delta_g}
E^{\mathrm{T}}_{QG}[\psi] 
= \frac{1}{2}\rho_o\int_M \psi \left(-\Delta_g\right) \psi \, dV_g  \ .
\eea

\medskip

\noindent {\bf Remarks}: 
\begin{itemize}
\item[a)] The geometric recast of $\hat{L}$, $E_{QG}$ and $E^{\mathrm{T}}_{QG}$
in terms of $\mathbf{g}$ does not require $N$ and $\tilde{\rho}$ (and $f_o$) to depend
only on $z$. Expressions (\ref{e:L_Delta_g}), (\ref{e:energy_gradient})
and (\ref{e:total_energy_Delta_g}) still hold for general functions $N=N(x, y,z)$ and 
$\tilde{\rho}=\tilde{\rho}(x, y,z)$
beyond the QG model. Such a freedom could be of interest,  e.g. in the study
of related perturbation and stability issues.

\item[b)] Denoting by $E^{\partial M}_{QG}$ the boundary term
\bea
\label{e:Energy_QG_boundary}
E^{\partial M}_{QG}[\psi] \equiv (-\frac{1}{2}\rho_o) \int_{\partial M} \psi 
\partial_{\mathbf{s}} \psi \, dA_h \ ,
\eea
energies $E^{\mathrm{T}}_{QG}[\psi]$ and $E_{QG}[\psi]$
are related, through integration by parts, as in (\ref{e:relation_Energies})
\bea
\label{e:relation_Energies_QG}
E^{\mathrm{T}}_{QG}[\psi] = E_{QG}[\psi] + E^{\partial M}_{QG}[\psi] \ .
\eea
As shown in the relation (\ref{e:QGenergy_1}), the QG energy 
$E_{QG}[\psi]$ has a kinetic component and an (available) potential energy one.
These are {\em bulk} contributions to the QG energy. If we consider $E^{\partial M}_{QG}[\psi]$
as a {\em boundary QG energy} contribution, we can write  
\bea
\label{e:relation_All_Energies_QG}
E^{\mathrm{T}}_{QG}[\psi] = E_{\mathrm{kin}}[\psi] + E_{\mathrm{AP}}[\psi] + E^{\partial M}_{QG}[\psi] \ ,
\eea
so that $E^{\mathrm{T}}_{QG}[\psi]$  accounts for the {\em total QG energy},
justifying the chosen notation. To gain physical intuition on the $E^{\partial M}_{QG}[\psi]$
term, let us consider a flat bottom profile, so that $\partial_{\mathbf{s}}=-\partial_z$
at the bottom. On the other hand, pressure and density anomalies are
respectively given by $p'= \rho_of_o\psi$ and $\rho'=-\frac{\rho_of_o}{g}\partial_z\psi$
(e.g. \cite{cushman2011}). This leads to
\bea
E^{\partial M}_{QG}[\psi] =  -\frac{g}{\rho_of^2_o}\int_{\partial M} p'\rho' dA_h \ .
\eea
\item[c)] As an academic remark, if we formally consider 
a negative $N^2<0$ (convection), the signature of the metric $\mathbf{g}$ changes 
from Riemannian to Lorentzian. This would change radically the properties of the metric, 
in particular involving the propagation of perturbations along null-cones
of angle $\arctan(|N|/f_o)$ from the vertical. This is however nonphysical
in the QG context, that assumes stable stratification $N^2>0$.
\end{itemize}
Focusing on the QG case, $N=N(z)$, the QG metric presents the following 
important geometric feature, namely that angles are not changed from the
flat Euclidean case.

\medskip

\noindent{\bf Lemma 2} (conformal flatness). {\em If $N=N(z)>0$ the QG
metric is conformally flat. In particular, with the change of variables 
$dz=\frac{f_o}{N(z)}d\zeta$, the QG metric (\ref{e:QG_effective_metric})
is written as
\bea
\label{e:QG_effective_metric_conf}
ds^2 = \left(\frac{\tilde{\rho}^2}{\rho_o^2}\right)\left(\frac{f_o^2}{N^2}\right)
\left(dx^2 + dy^2 + d\zeta^2\right) \ .
\eea
}
{\em Proof:} It suffices to note that for a non-vanishing $N=N(z)$, the relation
$dz=\frac{f_o}{N(z)}d\zeta$ in a domain not including the Equator (where $f_o=0$)
defines a good change of variables. Then the expression (\ref{e:QG_effective_metric_conf})
follows straightforwardly from (\ref{e:QG_effective_metric}). $\square$

\medskip

\noindent {\bf Remark}: 

\noindent The form (\ref{e:QG_effective_metric_conf}) makes apparent the 
conformally-flat character of the QG metric in the $N=N(z)$ case.
Other coordinates can be preferred in 
applications, making  convenient to have a coordinate-independent criterion characterizing 
conformal flatness [therefore testing $N=N(z)$].
In the present $3$-dimensional case, this is provided by the vanishing of 
the Cotton tensor or, equivalently the Cotton-York tensor,
defined in (\ref{e:Cotton}) and (\ref{e:Cotton_York}).

\subsection{An ocean drum}
\label{s:AnOceanDrum}
The expression of $\hat{L}$ and $E^{\mathrm{T}}_{QG}$ in terms of $\Delta_g$ makes natural the
study of
the eigenvalue problem (\ref{e:spectral_Delta})
for the QG metric (\ref{e:QG_effective_metric}).
The associated eigenfunctions can be chosen to be real ($\Delta_g$ is self-adjoint).
Introducing a constant $\alpha_o$ with the dimensions of a geostrophic 
streamfunction, i.e. $[\alpha_o]=(\mathrm{Length})^2(\mathrm{Time})^{-1}$, we 
can normalize these eigenfunctions to provide a natural orthogonal basis $\{\psi_n\}$
for the QG streamfunctions.

\medskip

\noindent{\bf Definition (Neumann ocean drum)}. {\em Given the QG
metric (\ref{e:QG_effective_metric}), we define a set of (real) eigenfunctions
$\{\psi_n\}$ of the associated (Neumann) Laplacian eigenvalue problem}
\bea
\label{e:spectral_Delta_QG}
-\Delta_g \psi_n = \lambda_n \psi_n  \ \ \ , \ \ \ 
\left.\partial_{\mathbf{s}} \psi_n \right|_{\partial M} = 0 \ ,
\eea
{\em with normalization given by}
\bea
\label{e:oceandrum_normalization}
\langle \psi_m | \psi_n \rangle_g = \int_M \psi_m \psi_n dV_g = \left(\alpha_o^2V_g\right) 
\delta_{nm} \ .
\eea
{\em The $\{\psi_n\}$ Laplacian eigenfunctions and the corresponding 
eigenvalues  $\lambda_n$ will be referred to as {\em ocean drum modes}
of the basin $M$.}

The ocean drum modes $\{\psi_n\}$, with dimensions of QG streamfunctions,
constitute an orthogonal basis in $L^2(M, \mu_g)$. Any streamfunction $\psi$
in $L^2(M, \mu_g)$, in particular
satisfying (\ref{e:vertical_BC}) but 
not necessarily Neumann boundary conditions, is uniquely written as
\bea
\label{e:psi_decomposition}
\psi = \sum_n^\infty C_n \psi_n \ \ , \ \ \hbox{with} \ C_n = 
\langle \psi | \psi_n \rangle_g =
\frac{1}{\alpha_o^2 V_g}\int_M \psi \, \psi_n \, dV_g \ ,
\eea
with $C_n$ dimensionless numbers. The norm of $\psi$ can be expressed as
\bea
\label{e:normalization_psi}
||\psi||_g^2 =\langle\psi|\psi\rangle_g =  \left(\alpha_o^2V_g\right) \sum_n |C_n|^2 \ .
\eea

\noindent {\bf Remark} (mixed ocean drum): 

\noindent The formulated (Neumann) ocean drum focuses on 
the vertical boundary condition (\ref{e:vertical_BC}), 
aiming at contacting QG baroclinic modes (cf. Lemma 3 below). 
A more general ocean drum  incorporating lateral boundary conditions
can be formulated in terms of a mixed eigenvalue problem \cite{Chavel84}: decomposing $\partial M$
in bottom-surface and lateral components as
$\partial M = (\partial M)_{\mathrm{BS}} \cup (\partial M)_{\mathrm{L}}$, set
$\left.\partial_{\mathbf{s}} \psi_n \right|_{ (\partial M)_{\mathrm{BS}}} = 0$ and 
$\left.\psi_n \right|_{(\partial M)_{\mathrm{L}}} = 0$. The mathematical essential 
ingredient is the homogeneous nature of the conditions, that guarantees
the self-adjointness of $(-\Delta_g)$ and completeness of $\{\psi_n\}$. Physically, it is remarkable that they also recover 
the spatially constant Dirichlet conditions $\left.\psi_n \right|_{(\partial M)_{\mathrm{L}}} = C(t)$ 
for lateral conditions in (\ref{e:QG_dynamics}) (indeed homogeneous up to a 
global constant, at each $t$). We are here interested in ocean basins where the area of the
bottom boundary dominates over the lateral one. As an approximation, in the following we will focus
on the Neumann ocean drum, which renders easier certain technical aspects
(in particular the discussion of heat kernel coefficients in \ref{s:HK_coefficients}).
A refined ocean drum is however possible.

\medskip

The following lemma
endows ocean drum modes $\{\psi_n\}$ with a physical content,
as a stationary basis (in the precise sense to be specified below)
onto which the dynamics and energetics of general QG
flows can be decomposed.

\medskip

\noindent{\bf Lemma 3 (stationary ocean drum modes)}. {\em The eigenfunctions $\{\psi_n\}$,
with $\{\lambda_n\}$ eigen-values, provide a stationary basis for QG
dynamics and energetics in the following sense:
\begin{itemize}
\item[i)] {\em Dynamics:} Eigenfunctions $\{\psi_n\}$ satisfy the QG
equations in the $f$-plane, i.e. for $f=f_o$ in the QG dynamical equation
(\ref{e:potential_vorticity}). In addition, for constant
$N$ and $\tilde{\rho}$ and for a flat bottom, the ocean
drum modes reduce to standard baroclinic modes.

\item[ii)] {\em Energetics:} Defining the fundamental energies $E_n$ as the QG
energies $E_{QG}[\psi_n]$ associated with ocean drum modes $\psi_n$, we have (setting $M_o\equiv\rho_oV_g$)
\bea
\label{e:energy_eigenvalues}
E_n \equiv E_{QG}[\psi_n] =  E^{\mathrm{T}}_{QG}[\psi_n] = \frac{1}{2} M_o\alpha_o^2\lambda_n \ .
\eea
Then, for sufficiently regular streamfunctions $\psi=\sum_n C_n \psi_n$ 
(namely, such that
the series converges in $L^2(M, \mu_g)$, but also in $H^1(M)$) not necessarily
satisfying $\left.\partial_{\mathbf{s}} \psi \right|_{\partial M} = 0$, 
the bulk QG energy $E_{QG}[\psi]$ is decomposed in ocean drum modes $E_n$ as
\bea
\label{e:Energy_modes}
E_{QG}[\psi] = \sum_n |C_n|^2 E_n \ ,
\eea
so that the total QG energy is 
\bea
\label{e:Energy_modes_total}
E^{\mathrm{T}}_{QG}[\psi] = \sum_n |C_n|^2 E_n +
(-\frac{1}{2}\rho_o) \int_{\partial M} \psi \partial_{\mathbf{s}} \psi \, dA_h \ .
\eea
\end{itemize}
}

\noindent{\em Proof:} 
Regarding point {\em i)} we note first that by construction, since we assumed $N\neq N(t)$ and 
$\tilde{\rho}\neq \tilde{\rho}(t)$,  modes $\psi_n$ are time independent, i.e. 
$\partial_t \psi_n = 0$. On the other hand, for $\psi=\psi_n$ (in the
f-plane  $f=f_o$) we have $q_n = \hat{L}\psi_n + f =
-\lambda_n \left(\frac{\tilde{\rho}^2}{\rho_o^2}\right)\left(\frac{f_o^2}{N^2}\right)\psi_n + f_o$,
where we have used relations (\ref{e:L_Delta_g}) and (\ref{e:spectral_Delta_QG}).
Then $\partial_t q_n = 0$ and, given that  $\tilde{\rho}$ and $N$ are only functions of $z$,
it holds $J(\psi_n, q_n)= -\lambda_n \left(\frac{\tilde{\rho}^2}{\rho_o^2}\right)
\left(\frac{f_o^2}{N^2}\right)J(\psi_n,\psi_n)=0$. Therefore (\ref{e:QG_dynamics_2}) is satisfied
and $\psi_n$ defines a stationary QG flow.
On the other hand, baroclinic modes are eigenfunctions of (the vertical-stretching part of)
Charney's operator with boundary conditions given by $\partial_z\psi|_{\partial M}=0$, consistent with the
vanishing vertical velocity ($w=0$) in the general QG boundary condition (\ref{e:vertical_BC}).
In the case of constant $N$ and $\tilde{\rho}$,
relation (\ref{e:L_Delta_g}) implies $-\Delta_g = \mathrm{const}\cdot \hat{L}$. In addition,
for a flat bottom the outgoing normal at the bottom is $\mathbf{s}=-\partial_z$, so that 
the Neumann condition (\ref{e:spectral_Delta_QG}) translates into $\partial_z\psi|_{\partial M}=0$. Then
Laplacian Neumann eigenmodes $\psi_n$ in (\ref{e:spectral_Delta_QG}) recover the
baroclinic eigenfunctions.

Regarding {\em ii)},  $E_{QG}[\psi_n] =  E^{\mathrm{T}}_{QG}[\psi_n]$ follows
from the Neumann boundary conditions for $\psi_n$ and relations (\ref{e:relation_Energies}).
Expression of $E_n$ in terms of $\lambda_n$ 
follows from (\ref{e:total_energy_Delta_g})
and the $\psi_n$ normalization. To prove (\ref{e:Energy_modes})
for a streamfunction $\psi$ not necessarily satisfying $\partial_{\mathbf{s}}\psi=0$ at the bottom,
we note that the assumption of convergence of the series $\psi=\sum_n C_n \psi_n$
in the $H^1$ norm $||\, ||_1$ (namely in $|| \psi ||^2_1= || \psi ||^2_{L^2} +
|| \mathbf{\nabla}\psi\cdot\mathbf{\nabla}\psi ||^2_{L^2}$), implies in particular
\bea
\label{e:lim_Sobolev}
\lim_{N\to\infty} \int_M \left(\nabla_i \psi - \sum_n^N C_n\nabla_i\psi_n\right)
\left(\nabla^i \psi - \sum_m^N C_m\nabla^i\psi_m\right)dV_g = 0 \ .
\eea
Then, for a given finite $N$ we can write
\bea
&&\int_M \left(\nabla_i \psi - \sum_n^N C_n\nabla_i\psi_n\right)
\left(\nabla^i\psi - \sum_n^N C_m\nabla^i\psi_m\right)dV_g  \nn \\
&=&
\int_M \nabla_i\psi \nabla^i \psi dV_g -2 \sum_n^N C_n\int_M \nabla_i\psi\nabla^i\psi_n dV_g
+ \sum_n^N\sum_m^N C_nC_m \int_M \nabla_i\psi_n \nabla^i \psi_m dV_g \nn\\
&=&
\int_M \nabla_i\psi \nabla^i \psi dV_g +2 \sum_n^N C_n\int_M \psi \Delta_g \psi_n dV_g
- \sum_n^N\sum_m^N C_nC_m \int_M \psi_n \Delta_g \psi_m dV_g \nn \\
&=&
\int_M \nabla_i\psi \nabla^i \psi dV_g -  \sum_n^N |C_n|^2 (\alpha^2_oV_g) \lambda_n \ ,
\eea
where in the second step we have integrated by parts
and used the Neumann conditions on $\psi_n$, and the third step
uses the eigenvalue relation (\ref{e:spectral_Delta})  and 
normalization (\ref{e:oceandrum_normalization}). Taking the 
limit $N\to\infty$,
using (\ref{e:lim_Sobolev}) and multiplying by $\frac{1}{2}\rho_o$, 
we recover the expression (\ref{e:Energy_modes}) for $E_{QG}[\psi]$. 
Expression for $E^{\mathrm{T}}_{QG}[\psi]$ in (\ref{e:Energy_modes_total})
follows from (\ref{e:Energy_modes}) and relation (\ref{e:relation_Energies}). $\square$
\medskip

\noindent {\bf Remarks}:
\begin{itemize}
\item[a)] The Laplace operator is guaranteed to be selfadjoint
for homogeneous Neumann ($\left.\partial_{\mathbf{s}} \psi_n \right|_{\partial M} = 0$), 
Dirichlet ($\left. \psi_n \right|_{\partial M} = 0$), mixed  
or, more generally, Robin boundary conditions [$\left(a\psi_n
+ b \,\partial_{\mathbf{s}} \psi_n \right)|_{\partial M} = 0$]. The  choice of Neumann ones
for the  ocean drum basis 
is justified by the recovery of baroclinic modes in point
{\em i)} of Lemma 3.

\item[b)] The bulk contribution to the energy is positive, $E_{QG} [\psi]\geq 0$.
The ocean drum eigenmodes have vanishing boundary energy contribution, 
$E^{\partial M}_{QG}[\psi_n]=0$, due to the Neumann boundary conditions.
In particular, it follows that ocean drum modes are all non-negative, $E_n\geq 0$
(in general, $\lambda_n\geq 0$ in the Laplacian eigenvalue problem with Neumann,
cf. (\ref{e:lambda_ordered}), 
Dirichlet or mixed boundary conditions).
However this does not exclude streamfunctions $\psi$ with 
negative total energy $E^{\mathrm{T}}_{QG}[\psi]<0$
(then $\left.\partial_{\mathbf{s}} \psi \right|_{\partial M} \neq 0$). 
This is accomplished, e.g., with eigenfunctions $\psi$ with negative eigenvalues.
The latter can be constructed with Robin conditions such that 
$a\cdot b <0$, if the boundary contribution in  (\ref{e:Energy_modes_total})
$E^{\partial M}_{QG}[\psi]= \frac{1}{2}\rho_o \int_{\partial_M} \left(\frac{a}{b}\right)
\psi^2 dA_h < 0$ is sufficiently large. Examples of such negative
eigenvalues of the Laplacian can be found in \cite{BerDen08}.

\end{itemize}

\subsubsection{Modified enstrophy.}
\label{s:modified_enstrophy}
The enstrophy invariant ${\cal E}[\psi]$ is defined in
(\ref{e:enstrophy_QG}) in terms of the square of $\hat{L}\psi$.
However its rewriting in terms of  $(-\Delta_g)$
is not natural from the spectral perspective of the ocean drum.
However, we can introduce a modified enstrophy
\bea
\label{e:modified_enstrophy_QG}
 \tilde{\cal E}[\psi] \equiv \int_M 
\left(\frac{\tilde{\rho}^2}{\rho_o^2}\right)^{-1}
\left(\frac{f_o^2}{N^2}\right)^{-1}\left(\hat{L}\psi\right)^2 \tilde{\rho}\,d^3x \ ,
\eea
also an invariant of the QG dynamics [since $\tilde{\rho}= \tilde{\rho}(z)$, $N=N(z)$], 
that is rewritten as 
\bea
\label{e:modified_enstrophy_QG_Delta} 
\tilde{\cal E}[\psi] = \rho_o \int_M \left(\Delta_g \psi\right)^2 dV_g \ ,
\eea
using (\ref{e:L_Delta_g}) and (\ref{e:determinant_g}).
This is the natural enstrophy invariant~\footnote{It is noteworthy
that, for the kinetic part, this is precisely the expression
obtained if using the metric $\mathbf{g}$ in the construction of the vector product
employed in the expressions of $\mathbf{v}_g$ and the relative vorticity,
instead of the flat one.} to be considered in 
a spectral discussion, since we have (for $\psi=\sum_n C_n \psi_n$ and up to boundary contributions)
\bea
\label{e:modified_enstrophy_QG_spectral} 
\tilde{\cal E}[\psi] = M_o \alpha_o^2 \sum_n |C_n|^2 \lambda_n^2 \ .
\eea

\section{Perspectives in ocean physics}
\label{s:OceanPhysics}
Hitherto, the essential elements in the discussion apply in both the 
oceanic and the atmospheric setting. In this section we focus on the ocean case.
In this context, the
assumption $\tilde{\rho}=\rho_o$ is a good approximation. 
The QG metric simplifies then to
\bea
\label{e:QG_metric_Ocean}
ds^2 = \frac{f_o^2}{N^2} \left(dx^2 + dy^2\right) + dz^2 =
\left(\frac{f_o^2}{N^2}\right) \left(dx^2 + dy^2 + d\zeta^2\right) \ ,
\eea
the second (explicitly conformally-flat) form assuming $N=N(z)$ (with $dz = \frac{f_o}{N}d\zeta$).
The effective QG metric is a geometric and physical property of an
(ocean mesoscale) mass of water, at a given latitude and with a given 
stratification characterised by the Brunt-V\"ais\"al\"a frequency $N$. 
In addition, the ocean drum modes $\{\psi_n\}$ and their
eigenvalues $\{\lambda_n\}$ are a property of a
given ocean basin. Such geometric recast can provide insight into relevant 
QG features, but it is not clear that
it will be particularly helpful in the explicit resolution of the 
QG equations. In contrast, it can prove 
useful in the study of generic and qualitative properties of QG flows,
in particular through the application of tools/concepts from
spectral geometry. This section discusses some points
in this spirit.

\subsection{Isotropy in geostrophic turbulence}
The isotropy assumption plays an important role in Charney's discussion of 
geostrophic turbulence \cite{Charney71}, in particular for deriving the equipartition
of energy between kinetic and available potential energy components.
Such hypothesis is justified from the structure
of the operator $\hat{L}$, together with reasonable physical assumptions.
In the present context, the conformally-flat form of the metric in (\ref{e:QG_metric_Ocean}) implies
that the (QG) ocean is effectively locally 
quasi-isotropic in scales over which 
the factor $\frac{f_o^2}{N^2}$ presents small
variations, so that it can be considered constant. 
The metric is then effectively the isotropic flat metric.
The Ricci scalar curvature $R$  of $\mathbf{g}$ provides a natural pointwise scale
for local variations of the geometry, through the  curvature radius 
defined as $1/L^2_{\mathrm{curvature}}\equiv |R|$.
This definition has the virtue of not depending on the 
coordinates.
Choosing $L_{\mathrm{isotropy}}\sim L_{\mathrm{curvature}}$ as the scale below which
the quasi-isotropic approximation holds, and using
\bea
\label{e:curvature_scalar}
R= -6\left(\partial_z\ln N\right)^2
+ 4\partial^2_z\ln N \ ,
\eea
following from (\ref{e:Ricci_scalar}) [or applying  
(\ref{e:Ricci_scalars_conf}) to the conformally-flat form of (\ref{e:QG_metric_Ocean})], we have
\bea
\label{e:L_isotropy}
L_{\mathrm{isotropy}} \sim \frac{1}{\left|6\left(\partial_z\ln N\right)^2
- 4\partial^2_z\ln N \right|^{\frac{1}{2}}} \ .
\eea
The isotropy hypothesis is well-established in the context
of geostrophic turbulence. In spite of this, the present geometric discussion can help
in providing a more systematic framework where physical 
assumptions are simplified and issues like the criticism in \cite{TunWel01}
can be naturally assessed. In this sense, it is worth noting that the adopted geometric approach
indicates the existence of privileged depths for isotropy (and therefore geostrophic
turbulence) characterized by the divergence of the scale of isotropy
$L_{\mathrm{isotropy}}\to \infty$.  
From the expression (\ref{e:L_isotropy}) this condition 
is fulfilled at depths satisfying 
\bea
\label{e:optimal_depths}
6\left(\partial_z\ln N\right)^2 = 4\partial^2_z\ln N \ .
\eea
This equation for $z$ defines {\em optimal QG depths} where 
isotropy is locally maximized.

\subsubsection{Aspect ratio.}
As a by-product, the expression of the {\em aspect ratio}
between typical vertical and horizontal scales, respectively
$\Delta z\sim L_{\mathrm{V}}$ and $\Delta x \sim L_{\mathrm{H}}$, does not depend 
on depth. This is apparent in the conformally-flat form (\ref{e:QG_metric_Ocean})
of the metric, from which
\bea
\frac{\Delta \zeta}{\Delta x} = 1 \ \ \Rightarrow \frac{L_{\mathrm{V}}}{L_{\mathrm{H}}}
\equiv \frac{\Delta z}{\Delta x} = \frac{f_o}{N} \ .
\eea
Of course, this is just the standard relation following from dimensional analysis estimations.
We simply point out that such a relation is actually ``exact'' in 
the QG model, without need of such estimations (at least, when separated from the boundaries).

\subsection{Deep ocean and hyperbolic space $\mathbb{H}^3$.}
\label{s:H3}
The typical vertical dependence of the Brunt-V\"ais\"al\"a frequency $N$ displays strong 
gradients in the upper layers of the ocean and then presents
a monotonic slow decay  at deeper layers, reaching comparatively 
small values at the bottom. More specifically, in the deep ocean
the $z$-dependence of $N$ 
is well approximated (e.g. \cite{PaoSwi12,LaCasce12}) by an exponential $N\sim N_o e^{-|z|/H}$ decay
(we assume here a negative $z$, with $z=0$ at the surface). 
We notice from (\ref{e:curvature_scalar}) that the Ricci curvature scalar then becomes a negative 
constant $R\sim -6/H^2$. This has an interesting geometric meaning, namely
indicating the hyperbolic nature of the QG metric at the
deep ocean. Integrating the relation $dz=(f_o/N(z))d\zeta$ 
with the exponential decay results in $N=\frac{f_o}{H}\zeta$.
Inserting this linear dependence in $\zeta$ in the conformally-flat
form of the QG metric (\ref{e:QG_metric_Ocean}), we find
\bea
\label{e:Poicare_half_plane}
ds^2  =g^{_{\mathbb{H}^3}}_{ij}dx^idx^j=\frac{H^2}{\zeta^2}\left(dx^2 + dy^2 + d\zeta^2\right) \ ,
\eea
which is the metric of the hyperbolic space $\mathbb{H}^3$
(in two dimensions it would correspond to the Poincar\'e half-plane).
In this specific sense $\mathbb{H}^3$, or more
precisely a compact patch of $\mathbb{H}^3$ with boundaries, provides a $0$-th order
approximation of the QG metric $g_{ij}$. 
Indeed
\bea
\label{e:conformal_hyperbolic}
g_{ij}= \Omega^2 g^{_{\mathbb{H}^3}}_{ij} =  \left(1 + o(\zeta)\right) g^{_{\mathbb{H}^3}}_{ij} \ ,
\eea
where $\Omega \equiv \frac{f_o}{N}\cdot \frac{\zeta}{H}$ and we have assumed
a dependence $N=\frac{f_o}{H}\zeta + o(\zeta^2)$ consistent
with the leading-order exponential decay of $N$ in $z$. Then,
defining $\tilde{\psi}\equiv \Omega^{-\frac{1}{2}}\psi$ and
using the conformal transformation properties of the Laplacian
in \ref{s:Conformal_transformations}, it holds
\bea
\langle \tilde{\psi} | (-\Delta_{g}) |\tilde{\psi}\rangle_{g}
= \langle \psi |  (-\Delta_{\mathbb{H}^3}^\Omega) | \psi \rangle_{\mathbb{H}^3} \ ,
\eea
where we have defined
\bea
\label{e:L_Omega}
\Delta_{\mathbb{H}^3}^\Omega = \Delta_{\mathbb{H}^3} - \left(\frac{1}{2}\Delta_{\mathbb{H}^3}\ln\Omega
+ \frac{1}{4}|\mathbf{\nabla} \ln\Omega|^2_{\mathbb{H}^3} \right) \ .
\eea
This selfadjoint operator (in the hyperbolic scalar product 
$\langle \cdot | \cdot \rangle_{\mathbb{H}^3}$) can be diagonalized 
\bea
-\Delta_{\mathbb{H}^3}^\Omega \psi^\Omega_n = \lambda^\Omega_n \psi^\Omega_n \ \ , \ \ 
\partial_{\mathbf{s}} \psi^\Omega_n|_{\partial M} = 0 \ ,
\eea
with $g^{_{\mathbb{H}^3}}_{ij} s^is^j=1$ and 
$\langle \psi^\Omega_n | \psi^\Omega_m \rangle_{\mathbb{H}^3} = \delta_{nm}$. 
Introducing $\tilde{\psi}^\Omega_n\equiv\Omega^{-\frac{1}{2}}\psi^\Omega_n$, it holds
\bea
\langle \tilde{\psi}^\Omega_n | (-\Delta_{g}) |\tilde{\psi}^\Omega_m\rangle_{g} = 
\lambda^\Omega_n \delta_{nm} \ .
\eea
Then, for small perturbations $\Omega \sim 1 + \epsilon \, \omega$  
of the metric around $\mathbb{H}^3$, we have
\bea
-\Delta_{\mathbb{H}^3}^\Omega = -\Delta_{\mathbb{H}^3} + \epsilon \frac{\Delta_{\mathbb{H}^3}\omega}{2} 
+ o(\epsilon^2) \ ,
\eea
and, to first order, the coefficients $\lambda^\Omega_n$ can be written as
\bea
\lambda^\Omega_n = \lambda^{\mathbb{H}^3}_n + \epsilon  
\frac{\langle \psi^{\mathbb{H}^3}_n |\frac{1}{2}\Delta_{\mathbb{H}^3}\omega|\psi^{\mathbb{H}^3}_n \rangle_{\mathbb{H}^3}}
{\langle \psi^{\mathbb{H}^3}_n | \psi^{\mathbb{H}^3}_n \rangle_{\mathbb{H}^3}} \ ,
\eea
where $\lambda^{\mathbb{H}^3}_n$ and $\psi^{\mathbb{H}^3}_n$ solve the hyperbolic
Laplacian eigenvalue problem
\bea
\label{e:Laplacian_eigenvalue_Hyperbolic}
-\Delta_{\mathbb{H}^3} \psi^{\mathbb{H}^3}_n = \lambda^{\mathbb{H}^3}_n \psi^{\mathbb{H}^3}_n 
\ \ , \ \ \partial_{\mathbf{s}} \psi^{\mathbb{H}^3}_n |_{\partial M} = 0 \ \ .
\eea
This motivates the systematic study of the Laplacian eigenvalue problem
(\ref{e:Laplacian_eigenvalue_Hyperbolic})
in compact submanifolds (with boundaries) of the hyperbolic space $\mathbb{H}^3$ (e.g. \cite{Bengu11}).

Finally, this hyperbolic leading-order behaviour
could be seen as indicating the existence of more (effective) 
volume available for QG degrees of freedom in the deep ocean 
than in the upper layers. Such hyperbolic nature of the QG deep ocean 
suggests geometric problems, such as: 
i) the ${\mathbb{H}^3}$ isoperimetric problem and
optimal forms in $\mathbb{H}^3$ 
providing qualitative descriptions of QG large
coherent structures (e.g. vortices), or ii) the role of (hyperbolically) 
separating geodesics in the study of QG instabilities.

\subsection{Large and small length scales: QG asymptotics and the heat kernel}
QG dynamics offers an appropriate description of the slow
motions of the ocean in the approximate length-scale range $L\sim20-200$ km. 
At larger or smaller scales the description of the flow must be
modified or complemented with other dynamical elements (limits depend on latitude and are not sharp, 
e.g. $L\sim50-500$ km also holds). 
In particular, planetary-geostrophic
equations account for large-scale ocean dynamics,
whereas at small scales QG dynamics makes a transition into the sub-mesoscale regime
requiring general primitive equations or even full Navier-Stokes for its description.
In this setting, a good qualitative control of low and large eigenvalues 
can provide insights into the mechanisms underlying 
the transition between these distinct dynamical regimes. 

\subsubsection{Large scales and small eigenvalues.} 
Low eigenvalues provide upper bounds for large QG scales.
The energy inverse-cascade 
of geostrophic turbulence transfers dynamically the energy injected at a given
scale into larger ones. But
energy cannot be stocked beyond a given largest scale $L^{\mathrm{QG}}_{\mathrm{max}}$ 
in QG modes, since the spectrum is
bounded below. In this context, taking $L^{\mathrm{QG}}_{\mathrm{max}}\sim 1/\sqrt{\lambda_1}$,
bounds in (\ref{e:bounds_lower_eigenvalues}) for the  
lowest eigenvalue $\lambda_1$ provide bounds (in particular an upper bound)
for such a largest scale, leading to
\bea
\label{e:larger_scale_QG}
(f_o/N)^{\frac{2}{3}}\left(L^2H\right)^{\frac{1}{3}} 
\sim V_g^{\frac{1}{3}} \lesssim L^{\mathrm{QG}}_{\mathrm{max}} \lesssim 2/h_c \sim 
(f_o/N)L^{\mathrm{basin}}_{\mathrm{max}}  
 \ ,
\eea
where $L$ and $H$ are, respectively, typical horizontal and vertical scales of the basin and
$L^{\mathrm{basin}}_{\mathrm{max}}$ is a  ``largest diameter'' of the latter. 
This particular estimate of the upper bound of $L^{\mathrm{QG}}_{\mathrm{max}}$
in terms of $L^{\mathrm{basin}}_{\mathrm{max}}$
 is a very crude one~\footnote{Note  that, given an actual physical 
basin, we can consider it divided
into smaller geometric sub-basins, and then study how the flow quantities defined 
in such sub-basins change according to 
different degrees of ``coarse graining'' of the flow. Analysis with such 
``renormalization'' flavour will be generically needed in the consistent definition of qualitative 
flow estimators  considered below in section \ref{s:diagnotic_calculation}.} (e.g. for mid-latitudes and a deep ocean,
we have  $f_o\sim 10^{-4}s^{-1}$ and $N\sim 10^{-3}s^{-1}$, so that $(f_o/N)\sim 10^{-1}$ and, 
with $L^{\mathrm{basin}}_{\mathrm{max}}\sim 5\cdot 10^3km$, we find 
$L^{\mathrm{QG}}_{\mathrm{max}}\sim 5\cdot 10^2km$).
However, finer studies (e.g. \cite{Simon83,AshBen93}) of  small eigenvalues ($\lambda_1, \lambda_2,...$) 
are of much interest for the statistical mechanics 
description of the largest scales of turbulent geophysical flows.
The reason is the condensation of energy in the first modes, a phenomenon 
occurring in  $2$-dimensional flows
as a consequence of the infinity of dynamical Casimir invariants in (\ref{e:Casimir}) 
(e.g. \cite{BouSim09,BouCor10,BouchetVenaille12}). 
When considering  only the conservation of energy-enstrophy, 
all energy is condensed in the first mode $\phi_1$. If more invariants are 
included, the energy is not limited to the first mode but is 
still concentrated in the first ones.
Low eigenvalues are then important when building criteria 
to assess qualitative properties of these flows, in particular
random changes in the topology of quasi-steady/equilibrium flow states, 
mainly controlled by the domain geometry (e.g. \cite{BouSim09,VenBou09,LoxNad13}).
In sum, this  $2$-dimensional-flow statistical mechanics setting  motivates the
systematic spectral geometry study of low eigenvalues of the $\Delta_g$ Laplacian,
with a particular focus on the assessment of the 
influence of stratification and topography
in the largest-scales dynamics [through the study of
curvature and boundary-value effects on ($\lambda_1, \lambda_2,...$)].
The $\mathbb{H}^3$ eigenvalue problem in section \ref{s:H3} can provide
an operational perturbative avenue to approximate expressions of such first eigenvalues.
In particular, we aim at applying this approach to low eigenvalues 
in the construction of invariant microcanonical
measures for  continuously stratified QG flows along the lines in \cite{BouCor10}.

In a more applied spirit, 
lowest eigenmodes can be heuristically used to construct diagnostic tools 
to assess the proximity of a QG flow (provided numerically or by
observational data) to a transition into a new non-QG dynamical regime
at large scales.  As an example, defining
the length scale $L$ of a flow 
with streamfunction $\psi$ as~\footnote{The factor $\frac{2}{d}=\frac{2}{3}$ 
is justified in \ref{s:partition_function}. Note also the alternative definition 
$L_{\mathrm{T}}^{-2} \equiv  \frac{2}{3} \cdot \frac{\langle \psi|(-\Delta_g)|\psi \rangle_g}
{\langle \psi|\psi \rangle_g}$, containing information on the behaviour at the basin
boundary.}
\bea
\label{e:L_scale_1st}
L^{-2} \equiv \frac{2}{3} \cdot
\frac{\int_M \nabla^i \psi \nabla_i \psi \; dV_g}{\langle \psi|\psi \rangle_g}
= \frac{2}{3}\cdot\frac{\sum_n \lambda_n |C_n|^2}{\sum_n |C_n|^2} \ ,
\eea
the Rayleigh quotient (\ref{e:Rayleigh_quotient}) indicates that, as long
as the following condition is satisfied
\bea
L^{2}\lambda_1 \leq \frac{3}{2} \ ,
\eea
there is ``room'' in the spectrum to accommodate the inverse-cascade flow of QG energy 
(note that $\lambda_1$ is a property of the basin, not of the flow).
Given a physical basin, the saturation of such condition 
($L^{2}\lambda_1 \sim \frac{3}{2}$) signals the transition
into a larger scale regime.

\subsubsection{Small scales and spectral asymptotics.} 
The small-scale behaviour is
controlled asymptotically by Weyl's law (\ref{e:Weyl_law}). Defining a 
wave number $k\equiv\sqrt{\lambda}$, a spectral density for large $k$ (per mode 
$\frac{dN(k)}{dk}$ and per mode and volume element $\frac{dN(k)}{dVdk}$)
is given by
\bea
\frac{dN(k)}{dk}\sim  3 \frac{B_3}{(2\pi)^3} V_g k^2 \ \ , \ \ 
\frac{dN(k)}{dVdk} \sim 3\frac{B_3}{(2\pi)^3} \frac{f^2_o}{N^2} k^2 \ ,
\eea
noting $dV_g=\frac{f^2_o}{N^2}dV$. This spectral density can then
be used in statistical mechanics studies of the energy spectrum at small scales,
by using it to relate $dE(k)/dk$ with the energy
per mode $dE/dN$. Rotation and stratification enter through the
factor $(f^2_o/N^2)$. In this context, corrections to the Weyl's law
asymptotic leading order coming from boundary and/or curvature terms \cite{Clark67,BalBlo70},
can play an important role in understanding the role of topography
and stratification in the energetics. In the following subsection
we consider this issue from a thermodynamical perspective, in the
context of a statistical mechanics toy-model  
based on the asymptotics to the heat kernel.

\subsubsection{Topography, stratification and a thermodynamical approach to ocean diagnosis.} 
\label{s:diagnotic_calculation}
When addressing the quantitative aspects of QG dynamics through the
resolution of the associated partial differential equations (PDE),
even a simplified model as the QG one requires a numerical treatment. 
This involves two kinds of problems: i) resolution  limitations  for sufficiently 
demanding physical settings (e.g. in global ocean models), 
and ii) efficient extraction of the relevant dynamical information 
in the resulting complex field configurations. 
Both problems, respectively of {\em a priori} and {\em a posteriori}
nature, demand an appropriate characterization of the qualitative
properties of the flow. The {\em a posteriori} issues also
apply for observational data of the mesoscale ocean. 
We consider now a tentative application of the
geometric approach complementary to the numerical PDE resolution,
with a specific interest in the diagnosis 
of stratification and topography effects in QG flows with structures of 
small typical scale, as compared to the basin size.

Specifically, we propose a family of thermodynamics-like
diagnostic estimators, derived from a statistical mechanics toy-model 
motivated by quasi-geostrophy (details are in \ref{s:statmech_toymodel}).
Such a model is not equivalent to the
full QG model (see \ref{s:Connection_QG_statmech}), but aims at
covering partial aspects of quasi-geostrophy,
with a special emphasis in the incorporation of topographic and stratification effects.
This is a methodological choice, namely a simplified analysis 
where some key QG elements are switched-off
(in particular, the crucial enstrophy constraint) in order to
isolate the addressed effects.
The main goal of such thermodynamical treatment is to define systematically,
from the thermodynamical potentials and derived quantities, a set of functionals 
of the streamfunction $\psi$ that can be 
used as diagnostics to characterize a given mass of water. 

\paragraph{Diagnostics calculation.} Given a QG $\psi$, the diagnostic estimators 
are constructed by first integrating the functions $1$, $|\psi|^2$ and 
$|\mathbf{\nabla}\psi|_g^2$ in the basin $M$. Specifically, we calculate
\bea
\label{e:diagnostics}
\int_M |\mathbf{\nabla}\psi|_g^2\, dV_g \ \ , \ \ V_g = \int_M dV_g  \ \ , \ \ \int_M |\psi|^2\, dV_g 
\ ,
\eea
respectively related to $(E, V_g, \mathrm{N})$, namely the energy, the (effective) basin volume and
the number of (quasi-)particles [see (\ref{e:N_n_toymodel})], characterizing
the thermodynamical macrostate 
associated with the statistical mechanics toy-model.
Although the diagnostic systematics 
is based on a thermodynamical formalism,
in its practical application we do not need to resort to that underlying justification.
We demonstrate here the approach in a particular example, and refer
to  \ref{s:statmech_toymodel} for further insight and details.

In a second step, an effective length-scale $L$ for the field $\psi$ 
is introduced by inverting
\bea
\label{e:E_L_text}
\!\!\!\!\!\!\!\!\!\!\!\! \frac{2}{3}\cdot \frac{\int_M |\nabla \psi|^2dV_g}{\int_M|\psi|^2dV_g} = \frac{1}{L^2}
\left(1 -
\frac{\sqrt{\pi}}{6} \frac{A_h}{V_g(L)} L - \frac{2}{3}\frac{a_1}{V_g(L)} L^2 
- \frac{b_1}{V_g(L)} L^3 + \dots \right) ,
\eea
with
\bea
\label{e:V_g_L}
V_g(L)\equiv \left(V_g +  \frac{\sqrt{\pi}}{2} A_h L + a_1 L^2 + b_1 L^3  
+ a_2 L^4 + \dots \right) \ .
\eea
This relation applies for flows with QG structures with a ``thermalized'' 
lengthscale $L$ much smaller than the basin scale, $L\ll (V_g)^{\frac{1}{3}}$. It 
generalizes the expression for $L$ in (\ref{e:L_scale_1st})
by taking into account stratification
and topography terms through the heat kernel coefficients $a_i$ and $b_i$.
The squared-length $L^2$ plays the role of an effective (inverse) temperature $T$, so that
expression (\ref{e:E_L_text}) is a $E=E(T, V_g,\mathrm{N})$ 
relation [see justification in (\ref{e:E_T}) and (\ref{e:E_L})].
The knowledge of $(L, V_g, \mathrm{N})$ permits to  write a (Helmholtz) free 
energy $F=F(L,V_g,\mathrm{N})$ from which thermodynamical quantities can be systematically
derived. In this example we give the expression for an effective pressure $P_g$
[cf. (\ref{e:PV})]
\bea
\label{e:PV_text}
\!\!\!\!\!\!\!\!\!\!\!\!P_g= 
\left(1 +  \frac{\sqrt{\pi}}{2} \frac{A_h}{V_g} L + \frac{a_1}{V_g} L^2 + \frac{b_1}{V_g} L^3  
+ \frac{a_2}{V_g} L^4 + \dots \right)^{-1}  \cdot\frac{\rho_o}{2L^2}\cdot
\frac{\int_M |\psi|^2 dV_g }{V_g} \ .
\eea
The interest of such a quantity $P_g$ is to provide a qualitative criterion for the
(mechanical) equilibrium between two masses of water $A$ and $B$: 
if  $(P_g)_A \neq (P_g)_B$, one could expect a reconfiguration of the
$\psi$ configuration tending to balance $(P_g)_A$ and $(P_g)_B$.
Similarly, other quantities can be constructed to monitor other types
of equilibria and, perhaps more interestingly, conditions for internal
stability [e.g. $C_V>0$ in (\ref{e:C_V})]. The violation of such stability conditions
could signal the need to incorporate new elements in the dynamical description,
in particular offering  diagnostic tools for the transition
from balanced to imbalanced motions (see e.g. \cite{McWilYavCul98,Nadiga14},
also \ref{s:stability}).

\subsubsection{Correlation functions.} Information of $\psi$ 
in the thermodynamical diagnostics above is only gained through (\ref{e:diagnostics}).
A finer set of observables, still for 
configurations with a ``thermal'' $L$, can be introduced 
through the expectation value for a field $\phi$
\bea
\label{e:expectation}
\langle \phi \rangle^L_g 
\equiv 
\frac{1}{\alpha^2_oV_g} \sum_i \langle \psi_i|\phi | \psi_i\rangle
e^{-L^2\lambda_i} \ .
\eea
In particular, if we consider expectation values of powers $\psi^n$ of the streamfunction $\psi$
\bea
\langle \psi^n \rangle^L_g 
=\frac{1}{\alpha^2_oV_g} \sum_i \langle \psi_i| \psi^n | \psi_i\rangle
e^{-L^2\lambda_i} \ ,
\eea
we can use  $\langle \psi^n \rangle_g^L$ to probe scales of the order
$\sim (L/n)$. In particular, still in a $L\ll (V_g)^{\frac{1}{3}}$
regime, we can use the heat kernel expansion to evaluate 
(\ref{e:expectation}) asymptotically, and construct estimators $\delta(\psi^n)$
aiming at probing the length scale at which stratification and topography effects 
become relevant. For this, we compare  $\langle \psi^n \rangle^L_g$ with
$\langle \psi^n \rangle^L_{\mathrm{flat}}$, the latter evaluated with a flat metric with  
$\bar{N}=\int_M N dV/\int_M dV$ (therefore
$V_{\mathrm{flat}}=\frac{f_o^2}{\bar{N}^2}V$, with $V$ the physical volume)
and a flat bottom. Using the coefficients in (\ref{e:HK_coefficients}) 
\bea
\!\!\!\!\!\!\!\!\!\!\!\!\!\!\!\!\!\!\!\!\!\!\!\!\!\!\!\!\delta_L(\psi^n) &\equiv& \frac{\langle \psi^n \rangle^L_g 
- \langle \psi^n \rangle^L_{\mathrm{flat}} }{\langle \psi^n \rangle^L_{\mathrm{flat}} } 
\sim \frac{b_1[\psi^n]}{a_0[\psi^n]} L + \frac{a_1[\psi^n]}{a_0[\psi^n]}  L^2+ 
 \frac{b_1[\psi^n]}{a_0[\psi^n]} L^3 
+  \frac{a_2[\psi^n]}{a_0[\psi^n]} L^4 + \dots  ,
\eea
using the approximation $a_0\sim(a_0)_{\mathrm{flat}}$. Such $\delta_L(\psi^n)$
can be used to probe
stratification and 
topography contributions, since the latter are encoded in $a_i[\psi^n]$
and $b_i[\psi^n]$.

\section{Conclusions and perspectives}
\label{s:Conclusions}
We have introduced an effective QG metric $\mathbf{g}$,  generically with non-vanishing curvature,
that constitutes a physical property of a QG mass of water associated with stratification.  
This permits to rewrite Charney's QG operator $\hat{L}$ in terms of
the associated scalar Laplace-Beltrami operator  $\Delta_g$ and, in particular, 
the (total) QG energy is given exactly by 
the expectation value of $(-\Delta_g)$ in the QG state defined by the streamfunction $\psi$. 
This has prompted the study of the
Laplacian, in particular its spectral properties, through the formulation of an ocean drum problem in the
ocean basin, namely the Neumann eigenvalue problem of $(-\Delta_g)$. The latter introduces a class of normal
modes $\psi_n$ whose vertical structure recovers standard QG barotropic/baroclinic modes for
constant Brunt-V\"ais\"al\"a frequencies and provides a f-plane stationary 
basis for a spectral decomposition of QG streamfunctions $\psi$. The corresponding eigenvalues 
$\lambda_n$ determine the length scale and energy of each mode. Although the exact determination of 
$\lambda_n$ is generically out of reach, spectral geometry tools can be used to estimate properties
of the spectrum in terms of curvature and boundary terms, accounting for stratification
and topography. This is the main contribution of the article, namely recasting QG elements
in a geometric form that can foster new avenues to QG dynamics complementary to existing ones.

In the specific application to the ocean context, we have used the conformally-flat nature of 
$\mathbf{g}$ to revisit the isotropy assumption in geostrophic turbulence, proposing the existence
of preferred depths with enhanced quasi-isotropy. This can be relevant in the
understanding of the vertical structure of mesoscale turbulence.
On the other hand, we have shown that the QG geometry of the deep ocean is well approximated 
by the hyperbolic metric in $\mathbb{H}^3$. We should generically expect average negative
values of the Ricci scalar curvature, sign changes signaling non-trivial configurations
of the QG field. This leads to a perturbative approach for studying 
deep ocean flows, with the hyperbolic space as the natural non-perturbed state.
In this setting, we have emphasized the interest of the hyperbolic
Laplacian eigenvalue problem in compact submanifolds of $\mathbb{H}^3$ with boundary.
Finally, we have discussed a class of global diagnostic quantities for numerical/observational data,
with a focus on stratification and topography effects. This is a tentative exploration
based on an ad hoc but systematic thermodynamical treatment, whose ultimate interest must be 
assessed from its usefulness as a tool.

Regarding next research steps, we are particularly interested in the application of the spectral ocean drum
problem to the  statistical mechanics study of geophysical flows, in
particular using RSM theory. A specific goal will be the explicit construction of an
invariant microcanonical measure for the continuously stratified QG model, by adapting the 
work in \cite{BouCor10}. Of special interest will be the study of stratification and topography 
through their incorporation into the spectral properties of the curved effective-metric Laplacian. 
A second line of research will be the extension of the presented geometric approach
to QG dynamics in order to incorporate fast/slow mode coupling in the setting
of the  wave-mean flow theory for geophysical fluids.
In particular, we will focus on the coupling between near-inertial internal waves and QG flows
in the Young-Ben Jelloul model \cite{YouBenJel97} and its refinement 
incorporating the back-reaction onto the QG field \cite{XieVan15}.


\medskip

\smallskip\noindent\emph{Acknowledgments.~} 

\noindent It is a pleasure to thank X. Carton, A. Colin de Verdi\`ere and R. Scott for their
scientific advice, patient guide and support. 
I thank all members of LPO for the warm atmosphere and particularly the scientific discussions with 
B. Blanke, T. Capuano, D. Ciani, D. Ferjani, A. Hochet, T. Huck, Q. Jamet, 
M. Kersal\'e, B. Lecann, 
P. L'Hegaret,  C. M\'enesguen, A. Ponte, 
G. Roullet, R. Schopp, A.M. Tr\'eguier and C. Vic.
I also thank F. Bouchet, T. David,  E. Gourgoulhon, B. Legras, T. Levasseur,
P.A. Luque, J.P. Nicolas, R.T. Pierrehumbert,
J. Vanneste, A. Venaille and B. Young.
Finally, I would like to express my deep debt of gratitude towards Sergio Dain.
This work is dedicated to his memory.

\appendix

\section{Coordinate expressions of covariant derivatives and curvature}
\label{s:Appendix_coordinateexpressions}
We complement section 
\ref{s:RiemannGeometry} with some more explicit expressions for covariant
derivatives and curvature.
Given a local chart $\{x^{i_1},\dots,x^{i_d}\}$, the 
associated basis for the tangent space (bundle) is given
by $\{\partial_{x^{i_1}},\dots,\partial_{x^{i_d}}\}$, whereas
$\{dx^{i_1},\dots,dx^{i_d}\}$ provides a basis for 
the cotangent bundle. A ${m \choose n}$ tensor field $\mathbf{T}$ can then be written as
$\mathbf{T} = {T^{{j_1}\dots {j_m}}}_{{i_1}\dots {i_n}}\partial_{x^{j_1}}\otimes \dots\otimes\partial_{x^{j_m}} 
\otimes dx^{i_1}\otimes \dots\otimes dx^{i_n}$ (summation over indices is assumed). 
The covariant derivative of a scalar field $\phi$ is just
given by standard partial derivatives $\nabla_i \psi = \partial_{x^i}\phi$, whereas
the covariant derivative of a ${1 \choose 0}$ tensor (i.e. a vector field)   $\mathbf{V} = V^i\partial_{x^i}$
is a ${1 \choose 1}$  tensor $\mathbf{\nabla V}$
\bea
\label{e:nablaV}
\mathbf{\nabla V} = \nabla_i V^j \partial_{x^j}\otimes d{x^i} \ \ , \ \ \hbox{with} \ \ 
\nabla_i V^j = \partial_{x^i} V^j + \Gamma^j_{ik}V^k \ ,
\eea
with the Christoffel symbols $\Gamma^k_{ij}$ of the Levi-Civita connection 
associated with the metric $\mathbf{g}=g_{ij}dx^i\otimes dx^j$ (the latter, 
namely a symmetric non-degenerate ${0\choose 2}$ tensor) given by
\bea
\label{e:Christoffel}
\Gamma^k_{ij} = \frac{1}{2}g^{kl}\left(\partial_{x^i}g_{lj} + \partial_{x^j}g_{li}
- \partial_{x^l}g_{ij}\right) \ .
\eea
The coordinate components of the ${1 \choose 3}$ Riemann  tensor introduced
in section \ref{s:connectioncurvature} are
\bea
\label{e:Riemann_Christoffel}
{R^i}_{jkl} = \partial_{x^k} \Gamma^i_{jl} - \partial_{x^l} \Gamma^i_{jk}
+ \Gamma^i_{km} \Gamma^m_{jl} - \Gamma^i_{lm} \Gamma^m_{jk} \ .
\eea
The Ricci curvature tensor is a symmetric ${0 \choose 2}$ tensor whose 
components are obtained by contracting indices in the Riemann tensor as $R_{ij}={R^k}_{ikj}$.
The Ricci scalar is obtained by first forming a ${1 \choose 1}$ tensor 
from the Ricci tensor by ``raising'' 
indices with the metric, i.e. ${R^i}_j=g^{ik}R_{kj}$, and then contracting
the remaining indices: $R={R^i}_i=g^{ij}R_{ij}$.

A property of $\mathbf{\nabla}$ playing a crucial role
in the geometric rewriting of Charney's operator in 
section \ref{s:QGRiemann}, 
is the fact that the divergence of a vector [trace of $\mathbf{\nabla V}$
in (\ref{e:nablaV})] can be expressed in terms of partial 
derivatives without the explicit appearance of the Christoffel symbols $\Gamma^k_{ij}$.
Using (\ref{e:Christoffel}) to rewrite $\Gamma^k_{ki}=\frac{1}{2}g^{kj}\partial_{x^i}g_{kj}
= \partial_{x^i}\ln\sqrt{g}$, it follows
\bea
\label{e:divervenceV}
\nabla_i V^i = \partial_{x^i} V^i + \Gamma^i_{ik}V^k 
= \frac{1}{\sqrt{g}}\partial_{x^i}\left(\sqrt{g}\;V^i\right) \ .
\eea
The Laplacian of a scalar field $\phi$ is introduced 
[Eq. (\ref{e:Laplacian_g})]
as the divergence of its gradient, 
$\Delta_g \phi = \nabla_i \nabla^i\phi$.
Writing the components of the gradient of $\phi$ as $\nabla^i\phi=g^{ij}\nabla_j\phi
= g^{ij}\partial_{x^j}\phi$  and
applying (\ref{e:divervenceV}) to $\nabla^i\phi$, it follows 
the expression (\ref{e:Laplacian_g_partial}) for $\Delta_g \phi$. Likewise for $K=\nabla_i s^i$.

\section{Heat kernel coefficients for Neumann boundary conditions}
\label{s:HK_coefficients}
Given the Laplacian eigenvalue problem (\ref{e:spectral_Delta}) with Neumann
homogeneous boundary conditions and given a scalar $\phi\in C^\infty(M)$,
it holds \cite{BransonGilkey90,Gilkey2003} (for normalized $\langle\tilde{\psi}_i|\tilde{\psi}_j\rangle_g
=\delta_{ij}$) 
\bea
\!\!\!\!\!\!\!\!\!\!\!\!\!\!\!\!\!\!\!\!\!\!\!\!\!\!\!\!\!\!\!\!
\mathrm{Tr}_{L^2}(\phi e^{-t(-\Delta_g)}) = \sum_i \langle \tilde{\psi}_i | \phi |\tilde{\psi}_i\rangle_g
e^{-t\lambda_i} \sim 
\frac{1}{(4\pi t)^{\frac{d}{2}}} \left(\sum_{n=0} a_n[\phi] t^n 
+ \sum_{n=0} b_n[\phi] t^\frac{2n+1}{2} \right)
\ \ (t\to 0) \ ,
\eea
with the first coefficients given by
\bea
\label{e:HK_coefficients}
a_0[\phi] &=& \int_M \phi \,dV_g \ ,  \\
b_0[\phi]&=& \frac{\sqrt{\pi}}{2} \int_{\partial M} \phi\, dA_h \ , \nn \\
a_1[\phi] &=& \frac{1}{6}
\left(\int_M \phi R \, dV_g + \int_{\partial M}  (2\phi K - 3 \partial_{\mathbf{s}}\phi)  \, dA_h\right)  \ , \nn \\
b_1[\phi] &=& \frac{\sqrt{\pi}}{2} \frac{1}{96}
\int_{\partial M} \left[\phi\left(16 R + 8 R_{ij}s^is^j  + 13 K^2 + 2 K_{ij}K^{ij}  \right)
- 6 \partial_{\mathbf{s}}\phi K + 24\partial_{\mathbf{s}}\partial_{\mathbf{s}}\phi \right] dA_h \nn  \ , \\
a_2[\phi] &=&  \frac{1}{360}
\left[\int_M \phi\left( 12 \nabla^i\nabla_i R + 5 R^2
-2  R_{ij}R^{ij} + 2  R_{ijkl}R^{ijkl} \right)dV_g \right. \nn \\
&&\left.+ \int_{\partial M}\phi\left(-42 \partial_{\mathbf{s}} R + 24 D^iD_i K 
+ 20 R K + 4 R_{ij}s^is^j K \right.\right.\nn \\
&&\left.\left. -12  R_{ikjl}s^ks^l K^{ij} + 4  R_{ij} K^{ij}
+ \frac{40}{3}K^3 + 8 K_{ij}K^{ij} K + \frac{32}{3} {K^i}_j{K^j}_k{K^k}_i \right.  \right.\nn \\
&&\left.\left. -\partial_{\mathbf{s}}\phi \left( 30 R 
+ 12 K^2 +12 K_{ij}K^{ij} \right)
+ 24 \partial_{\mathbf{s}}\partial_{\mathbf{s}}\phi K -30 \partial_{\mathbf{s}}\nabla^i\nabla_i\phi
 \right)dA_h 
\right]\nn  \ .
\eea
Heat kernel asymptotics are recovered with $\phi=1$. Denoting 
$a_i[1]=a_i$ and $b_i[1]=b_i$
\bea
\label{e:HK_expansion}
K(t)= \sum_i  e^{-t\lambda_i} \sim \frac{1}{(4\pi t)^{\frac{d}{2}}} \left(\sum_{n=0} a_n t^n + \sum_{n=0} b_n t^\frac{2n+1}{2} \right)
\ \ (t\to 0) \ .
\eea

\section{Curvature elements of the quasi-geostrophic metric}
\label{s:QG_geometry}
We collect some curvature elements of the QG metric
(\ref{e:QG_metric_Ocean}), needed in (\ref{e:HK_coefficients}). 
The Ricci scalar, the square of the Ricci tensor and
the Kretschmann scalar are, respectively
\bea
\label{e:Ricci_scalar}
R= -\frac{2}{N^{2}}{\left(5 \, \left(\frac{\partial\,N}{\partial z}\right)^{2} - 2 \, N \frac{\partial^2\,N}{\partial z^2}\right)}  \ , \\
R_{ij}R^{ij}=\frac{2}{N^{4}} \, {\left(17 \, \left(\frac{\partial\,N}{\partial z}\right)^{4} - 14 \, N \left(\frac{\partial\,N}{\partial z}\right)^{2} \frac{\partial^2\,N}{\partial z^2} + 3 \, N^{2} \frac{\partial^2\,N}{\partial z^2}^{2}\right)}  \ , \\
R_{ijkl}R^{ijkl}=\frac{4}{N^{4}} \, {\left(9 \, \left(\frac{\partial\,N}{\partial z}\right)^{4} - 8 \, N \left(\frac{\partial\,N}{\partial z}\right)^{2} \frac{\partial^2\,N}{\partial z^2} + 2 \, N^{2} \frac{\partial^2\,N}{\partial z^2}^{2}\right)} \ ,
\eea
where $N=N(z)$ is assumed. We note that they are
independent of $f_o$, depending only on $N$.
 It is interesting to remark that a dependence in the
latitude enters if we relax the QG stratification to
a general $N=N(x,y,z)$. We have then,  e.g. for the Ricci scalar
\bea
\!\!\!\!\!\!\!\!\!\!\!\!\!\!\!\!\!\!\!\!\!\!\!\!\!\!\!\!\!\!\!\!\!\!\!\!R =-\frac{2 \, {\left(N^{2} \left(\frac{\partial\,N}{\partial x}\right)^{2} - N^{3} \frac{\partial^2\,N}{\partial x^2} + N^{2} \left(\frac{\partial\,N}{\partial y}\right)^{2} - N^{3} \frac{\partial^2\,N}{\partial y^2} + 5 \, f_o^{2} \left(\frac{\partial\,N}{\partial z}\right)^{2} - 2 \, f_o^{2} N\frac{\partial^2\,N}{\partial z^2}\right)}}{f_o^{2} N^{2}} \ .
\eea
Regarding the extrinsic metric terms of the bottom boundary in $\partial{M}$, let us consider a
topography profile $\eta=\eta(x,y)$. Then, the (outgoing) normal vector is
\bea
\!\!\!\!\!\!\!\!\mathbf{s}=\frac{1}{\sqrt{{\left(\left(\frac{\partial\,\eta}{\partial x}\right)^{2} + \left(\frac{\partial\,\eta}{\partial y}\right)^{2}\right)} \frac{N^{2}}{f_o^{2}} + 1}}\left(\frac{N^{2}}{f_o^{2}} \cdot \frac{\partial\,\eta}{\partial x} \,\frac{\partial}{\partial x } + \frac{N^{2}}{f_o^{2}}\cdot\frac{\partial\,\eta}{\partial y}\,\frac{\partial}{\partial y } -\frac{\partial}{\partial z }\right) \ ,
\eea
from where it follows
\bea
\partial_\mathbf{s}R=-\frac{4 \, {\left(5 \, f_o \left(\frac{\partial\,N}{\partial z}\right)^{3} - 6 \, f_o N \frac{\partial\,N}{\partial z} \frac{\partial^2\,N}{\partial z^2} + f_o N^{2} \frac{\partial^3\,N}{\partial z^3}\right)}}{ N^{3}\sqrt{{\left(\left(\frac{\partial\,\eta}{\partial x}\right)^{2} + \left(\frac{\partial\,\eta}{\partial y}\right)^{2}\right)} N^{2} + f_o^{2}}} \ .
\eea
If we consider the corresponding extrinsic curvature $K_{ij}$, its trace $K$ has the form
\bea
\!\!\!\!\!\!K=\frac{-1}{f_o N \left({{\left(\frac{\partial\,\eta}{\partial x}^{2} + \frac{\partial\,\eta}{\partial y}^{2}\right)} N^{2} + f_o^{2}}\right)^{\frac{3}{2}}} 
\left[{\left(2 \, \frac{\partial\,\eta}{\partial x} \frac{\partial^2\,\eta}{\partial x\partial y} \frac{\partial\,\eta}{\partial y} - \frac{\partial^2\,\eta}{\partial x^2} \frac{\partial\,\eta}{\partial y}^{2} 
\frac{\partial\,\eta}{\partial x}^{2} \frac{\partial^2\,\eta}{\partial y^2}\right)} N^{5}
\right. \nn\\
\left. - {\left(f_o^{2} \frac{\partial^2\,\eta}{\partial x^2} + f_o^{2} \frac{\partial^2\,\eta}{\partial y^2}\right)} N^{3} - {\left(2 \, f_o^{4} + 3 \, {\left(f_o^{2} \frac{\partial\,\eta}{\partial x}^{2} + f_o^{2} \frac{\partial\,\eta}{\partial y}^{2}\right)} N^{2}\right)} \frac{\partial\,N}{\partial z}\right] \ .
\eea
Similar (but much longer expressions) hold for $K_{ij}K^{ij}$
and ${K^i}_j{K^j}_k{K^k}_i$.

\section{Conformal transformations}
\label{s:Conformal_transformations}
Let us consider two metrics $\mathbf{g}$ and  $\tilde{\mathbf{g}}$ on a $d$-dimensional
manifold $M$, conformally related as $\tilde{g}_{ij} = \Omega^2g_{ij}$.
Then, their 
Ricci scalar curvatures are related as (see e.g. \cite{Wald84})
\bea
\label{e:Ricci_scalars_conf}
\!\!\!\!\!\!\!\!\!\!\!\!\tilde{R} = \Omega^{-2}\left[R - 2(d-1)g^{ij}\nabla_i\nabla_j\ln\Omega
- (d-2)(d-1)g^{ij}\nabla_i\ln\Omega\nabla_j\ln\Omega\right] \ .
\eea
On the other hand, given a scalar $\phi$ on $M$ and defining a
conformally transformed scalar with conformal weight $s$ as 
$\tilde{\phi}=\Omega^s\phi$, it holds for the respective Laplacians
\bea
\Delta_{\tilde{g}}\tilde{\phi} &=&\Omega^{s-2}\Delta_{g}\phi
+(2s+d-2)\Omega^{s-3}g^{ij}\nabla_i\Omega\nabla_j\phi \nn \\
&&+ s \Omega^{s-3}\phi \Delta_{g}\Omega + s(d+s-3)\Omega^{s-4}\phi
g^{ij}\nabla_i\Omega\nabla_j\Omega \ .
\eea
In dimensions $d\geq4$, the vanishing of the 
Weyl tensor (traceless part of the Riemann tensor, e.g. \cite{Wald84}) 
characterizes a given metric as conformally flat.
However in dimension $d=3$ the  Weyl tensor identically vanishes.
Conformal flatness is then characterized
by the vanishing of the Cotton tensor that, in dimension $d$, is
given by 
\bea
\label{e:Cotton}
C_{ijk} = \nabla_k R_{ij} - \nabla_j R_{ik} 
+ \frac{1}{2(d-1)}\left(\nabla_jR\, g_{ik} - \nabla_kR\, g_{ij} \right) \ .
\eea
The same information is contained in the (Hodge-)dual Cotton-York tensor 
\bea
\label{e:Cotton_York}
{C_i}^j = \nabla_k \left(R_{li}-\frac{1}{4}Rg_{li} \right)\epsilon^{klj} \ .
\eea
Then $C_{ijk}$ and ${C_i}^j$ vanish for the conformally-flat QG metric (\ref{e:QG_effective_metric_conf})
(with $N=N(z)$).
Deviations from QG, $N=N(x,y,z)$, can be invariantly characterized by the
eigenvalues of the symmetric trace-free $C_{ij}$ or, equivalently,
by its {\em principal invariants} (characteristic polynomial coefficients):
$\iota_2=-\frac{1}{2}C_{ij}C^{ij}$ and $\iota_3=\frac{1}{3}{C^i}_j{C^j}_k{C^k}_i$
(since $\iota_1={C_i}^i=0$).

\section{A weakly-interacting-particle ``QG model''}
\label{s:statmech_toymodel}
Given a flow subject to QG dynamics, let us introduce the following 
``particle'' toy-model. Let us consider a
set of $\mathrm{N}$ identical systems, referred to 
as {\em elementary QG excitations} or {\em QG quasi-particles}, 
distributed over (stationary) energy levels $E_n$ given by (\ref{e:energy_eigenvalues}),
i.e. $E_i=\frac{1}{2}  M_o \alpha_o^2\lambda_i$ (with $M_o=\rho_oV_g$ a
constant in the model). For a given QG state
characterised by $\psi=\sum_i C_i \psi_i$, let us define 
the
{\em level-occupation numbers} 
$(n_1, n_2, \dots , n_i, \dots)$ and the {\em total number} ${\mathrm N}$ of
QG quasi-particles as
\bea
\label{e:N_n_toymodel}
{\mathrm N} \equiv\frac{\int_M|\psi|^2dV_g}{\alpha^2_oV_g}  = \frac{\langle\psi|\psi\rangle_g}{\alpha^2_oV_g}
\ \ \ , \ \ \ n_i \equiv |C_i|^2  \ ,
\eea
where $\alpha_o$ is chosen such that ${\mathrm N}\gg 1$ (the present discussion
does not need to fix $\alpha_o$). 
Then, denoting $E=E_{QG}[\psi]$ and using expression (\ref{e:Energy_modes})
and the norm (\ref{e:normalization_psi}) of $\psi$
\bea
\label{e:E_N_toymodel}
E = \sum_i n_i E_i \ \ , \ \ \mathrm{N} = \sum_i n_i \ .
\eea
These relations correspond to the distribution of $\mathrm{N}$ (identical)
particles with total energy $E$ among the energy levels $E_i$ of a {\em mono-particle}
(quantum) Hamiltonian. 
Expressions (\ref{e:E_N_toymodel}) are a reminiscent of 
the  (quantum) statistical mechanics treatment~\footnote{A key feature for writing (\ref{e:E_N_toymodel})
is the absence of normalization of $\psi$, a fundamental difference with quantum mechanics
that allows us to interpret the (squared) norm of $\psi$ as the total number of particles.} of the ideal gas, where
the monoparticle  Hamiltonian is proportional to the
Laplacian $(-\Delta_g)$~\footnote{Indeed, this analogy is motivated by the
identification \cite{HolmZeitlin98} of $E_{QG}[\psi]$ in (\ref{e:QGenergy_1}) 
as a QG classical (non-canonical) Hamiltonian, together with its rewriting 
 (\ref{e:total_energy_Delta_g}) as an exact Laplacian (up to
boundary terms).
Expressions (\ref{e:E_N_toymodel}) then
stand in the spirit of a formal Fock 
``second quantization'' of $\psi$ as a free field.}. 
Of course there are fundamental differences between a QG flow and the ideal gas, 
in particular regarding the existence of further constraints among the $E_i's$. Whereas for
weakly interacting particles (strictly non-interacting in
the ideal gas)  expressions (\ref{e:E_N_toymodel}) encode the whole information,
the infinite number of conserved quantities (\ref{e:Casimir}) in QG dynamics introduce further constraints. 
In particular, the  enstrophy conservation 
imposes ${\cal E}\sim\sum_i n_i  E_i^2$ [cf. expression
(\ref{e:modified_enstrophy_QG_spectral})]. 
This introduces non-local and non-linear interactions 
among the modes, crucial in geostrophic turbulence.
In the present particle toy-model we make the drastic assumption of neglecting such non-local
interactions.
 
\medskip
\noindent {\bf Definition (particle toy-model)}. 
{\em Given a basin with ocean-drum modes $\{\psi_i\}$ 
and a QG flow with $\psi=\sum_i C_i\psi_i$, we define a
(microcanonical ensemble) 
statistical mechanics model of $\mathrm{N}=\langle\psi|\psi\rangle_g/(\alpha^2_oV_g)$ 
``weakly-interacting'' identical particles,
with total energy $E=E_{QG}[\psi]$, distributed 
in the (mono-particle) ocean-drum levels $E_i$ with occupation numbers
$n_i=|C_i|^2$, such that  $E = \sum_i n_i E_i$, $\mathrm{N} = \sum_i n_i$ hold
without further constraints.}

\medskip
\noindent {\bf Remarks}:
\begin{itemize}
\item[a)] This is an {\em ad hoc}  statistical mechanics
  model motivated, but different, from 
QG dynamics. It will be ultimately justified if it
provides new insights into aspects of the QG
model.

\item[b)] Although interactions arising from constraints are eliminated, the model
still contains non-local interactions related to topography and stratification (see
below). Methodologically,
this offers an avenue to isolate and assess the effects of topography and stratification,
before ``switching on'' the rest of actual physical interactions.

\item[c)] The model adopts an equilibrium statistical mechanics treatment, a
further assumption to be assessed in different particular regimes of QG dynamics.

\end{itemize}

\subsection{Partition function and free energy: heat kernel expansion.}
\label{s:partition_function}
Systems with long-range interactions, such as $2$-dimensional fluids,
display subtle issues concerning statistical ensemble
equivalence (e.g. \cite{Campa200957,CamDauDau14,BouchetVenaille12}). In the present toy-model 
we shall relax such concerns and will
adopt a canonical ensemble treatment,
assuming equilibrium with a thermal bath with formal ``temperature'' $T$. 
Such temperature is not the (molecular agitation) physical
one, but rather characterizes an equilibrium
parameter of the (weakly-interacting) QG particles. As argued below,
it is related to a length scale $L$ in a regime of QG dynamics where 
coherent structures have approximately the same size $L$, much smaller than the basin
one. The canonical partition function $Z(T, V_g, \mathrm{N})$ of a system of $\mathrm{N}$ 
weakly-interacting identical particles can be written as 
\bea
Z(T, V_g, \mathrm{N}) = \frac{1}{\mathrm{N}!}Z_{\Delta_g}(T,V_g)^{\mathrm{N}} \ , \
\eea
with  $Z_{\Delta_g}(T,V_g)$ the mono-particle partition function 
\bea
Z_{\Delta_g}(T,V_g) = \sum_i e^{-\beta E_i} = \sum_i e^{-\frac{E_i}{k_{\mathrm{B}}T}} = 
\sum_i e^{-\frac{M_o\alpha_o^2}{2k_{\mathrm{B}}T}\lambda_i} \ ,
\eea
where $\beta =1/(k_{\mathrm{B}}T)$, with $k_{\mathrm{B}}$ the Boltzmann constant. 
Introducing the length scale $L$ (the relation with the length scale
introduced in (\ref{e:L_scale_1st}) will be 
clarified below)
\bea
\label{e:Length_scale_Stat}
L^2 \equiv \frac{M_o\alpha_o^2}{2k_{\mathrm{B}}T} \ ,
\eea
we have
\bea
Z_{\Delta_g}(L,V_g) =
\sum_i e^{-L^2 \lambda_i} \ \ , \ \ 
Z(L, V_g, \mathrm{N}) = \frac{1}{\mathrm{N}!}\left( \sum_i e^{-L^2 \lambda_i}\right)^{\mathrm{N}} \ \ . 
\eea
Hitherto the expression of $Z(T, V_g, \mathrm{N})$ is exact in our toy-model. 
To take a step further and express it in terms of geometric features of $\mathbf{g}$,
we can use the heat kernel expansion 
for small scales $L\to 0$ (namely $\frac{L}{V_g^{1/3}}\sim\frac{L}{L^{\mathrm{basin}}_{\mathrm{max}}} 
\ll 1$).
Making $t=L^2$ in (\ref{e:Heat_Kernel}) (or (\ref{e:HK_expansion}))
\bea
\label{e:partition function_L}
Z(L, V_g, \mathrm{N}) = \frac{1}{\mathrm{N}!}\frac{1}{(4\pi L^2)^{\frac{3\mathrm{N}}{2}}}
V_g(L)^{\mathrm{N}} \ ,
\eea
with
\bea
V_g(L)\equiv \left(V_g +  \frac{\sqrt{\pi}}{2} A_h L + a_1 L^2 + b_1 L^3  
+ a_2 L^4 + \dots \right) \ ,
\eea
where coefficients $a_i$ and $b_i$  are given in \ref{s:HK_coefficients}
(in (\ref{e:HK_coefficients}) with $\phi=1$).
Heat kernel coefficients encode curvature (in $a_i$ and $b_i$) and boundary terms (in $b_i$),
bringing stratification and topography interactions, respectively, into the partition
function.
The non-divergent part of the asymptotic series is analytic
in $L$. It is also convenient to express 
the expansion of $Z$ in terms
of the effective temperature $T$, using (\ref{e:Length_scale_Stat})
\bea
\label{e:partition function_T}
Z(T, V_g, \mathrm{N}) = \frac{1}{\mathrm{N}!}\left(\frac{k_{\mathrm{B}}T}{2\pi M_o\alpha_o^2}\right)^{\frac{3\mathrm{N}}{2}}
V_g(T)^{\mathrm{N}} \ ,
\eea
where now (note that the non-divergent part of the expansion is not analytic in $1/T$)
\bea
\!\!\!\!\!\!\!\!\!V_g(T)\equiv \left(V_g +  \frac{\sqrt{\pi}}{2} A_h 
\left(\frac{M_o\alpha_o^2}{2k_{\mathrm{B}}T}\right)^{\frac{1}{2}} 
+ a_1 \frac{M_o\alpha_o^2}{2k_{\mathrm{B}}T} + b_1 \left(\frac{M_o\alpha_o^2}{2k_{\mathrm{B}}T}\right)^{\frac{3}{2}} 
+ \dots \right) \ .
\eea
Since $\int_M |\psi|^2dV_g$ is not preserved by the QG flow, the grand-canonical ensemble $\Xi$ 
(where the number of particles is not preserved) is a natural statistical ensemble. It is given by
\bea
\label{e:grandcanonical}
\Xi(T, V_g, \mu) = \sum_{\mathrm{N}} \sum_i e^{\frac{-1}{k_{\mathrm{B}}T}\left(E_{\mathrm{N},i} - \mu \mathrm{N} \right)}
= \sum_{\mathrm{N}} e^{\frac{\mu \mathrm{N}}{k_{\mathrm{B}}T}} Z(T, V_g, \mathrm{N}) \ ,
\eea
with $E_{\mathrm{N},i}$ the energy of the state with $\mathrm{N}$ particles and
$\mu$ the {\em chemical potential}. We shall give a more explicit expression below
(for small $L$), using canonical ensemble elements.

\subsection{Thermodynamics elements.} The (Helmholtz) free energy $F\!\!=\!\!E\!-\!TS$ 
is built from the canonical partition function 
as
\bea
\label{e:F_GQ}
\!\!\!\!\!\!\!\!\!\!\!\!F = -k_{\mathrm{B}}T \ln Z(T, V_g, \mathrm{N}) = 
-\mathrm{N}k_{\mathrm{B}}T\left(\frac{3}{2}\ln T + \ln\left(\frac{V_g(T)}{\mathrm{N}}\right) + \mathrm{const}  \right) \ ,
\eea
from which thermodynamical quantities can be systematically derived. First derivatives 
\bea
S = -\left(\frac{\partial F}{\partial T}\right)_{(V_g,\mathrm{N})} \ , \
P_g = -\left(\frac{\partial F}{\partial V_g}\right)_{(T,\mathrm{N})} \ , \
\mu = \left(\frac{\partial F}{\partial N}\right)_{(T,\mathrm{N})} \ ,
\eea
provide the entropy $S$,  an effective pressure $P_g$ (conjugated to the effective volume $V_g$)
and the chemical potential $\mu$.
Heat capacities and compressibilities follow from
second derivatives of the thermodynamical potentials (e.g. \cite{Callen85}).
The expression of $P_g$ is 
\bea
\label{e:PV}
P_gV_g\left(1 +  \frac{\sqrt{\pi}}{2} \frac{A_h}{V_g} L + \frac{a_1}{V_g} L^2 + \frac{b_1}{V_g} L^3  
+ \frac{a_2}{V_g} L^4 + \dots \right) = \frac{\mathrm{N} M_o\alpha_o^2}{2L^2} \ ,
\eea
or, perhaps more transparently, in terms of $T$ 
\bea
\label{e:PV(T)}
\!\!\!\!\!\!\!\!\!\!\!\!\!\!\!\!\!\!\!\!\!\!\!\!\!\!\!\!\!\!\!\!\!\!\!
P_gV_g\left(1 +  \frac{\sqrt{\pi}}{2} \frac{A_h}{V_g} 
\left(\frac{M_o\alpha_o^2}{2k_{\mathrm{B}}T}\right)^{\frac{1}{2}} 
+ \frac{a_1}{V_g} \left(\frac{M_o\alpha_o^2}{2k_{\mathrm{B}}T}\right) 
+\frac{b_1}{V_g} \left(\frac{M_o\alpha_o^2}{2k_{\mathrm{B}}T}\right)^{\frac{3}{2}}+ \dots\right) &=& 
\mathrm{N}k_{\mathrm{B}}T
\ .
\eea
The corrections in $V_g=V_g(T)$ to the ideal gas equation of state indicate the 
presence of effective interactions due to stratification  and
topography, encoded in curvature and boundary terms in the
heat kernel coefficients (a real fluid modelled e.g. with Van der Waals equation of state
would also have $P_g=P_g(T)$ corrections).
It also follows for $\mu$ 
\bea
\label{e:mu_QG}
\mu = k_{\mathrm{B}}T\ln\left(\frac{\mathrm{N}}{V_g(T)}
\left(\frac{2\pi M_o\alpha_o^2}{k_{\mathrm{B}}T}\right)^{\frac{3}{2}}\right) \ .
\eea
The rest of thermodynamical quantities follow similarly.
Particularly important is the relation between the energy 
$E$ and temperature $T$ (or alternatively, the length scale $L$)
\bea
\label{e:E_T}
\!\!\!\!\!\!\!\!\!\!\!\!\!\!\!\!\!\!\!\!\!\!\!\!\!\!\!\!\!\!\!\!\!\!\!E&=&\langle E \rangle = - \frac{\partial}{\partial \beta}\ln Z \\
\!\!\!\!\!\!\!\!\!\!\!\!\!\!\!\!\!\!\!\!\!\!\!\!\!\!\!\!\!\!\!\!\!\!\!&=&\frac{3}{2}\mathrm{N}k_{\mathrm{B}}T\left(1 -
\frac{\sqrt{\pi}}{6} \frac{A_h}{V_g(T)} 
\left(\frac{M_o\alpha_o^2}{2k_{\mathrm{B}}T}\right)^{\frac{1}{2}} 
- \frac{2}{3}\frac{a_1}{V_g(T)} \left(\frac{M_o\alpha_o^2}{2k_{\mathrm{B}}T}\right) 
- \frac{b_1}{V_g(T)} 
\left(\frac{M_o\alpha_o^2}{2k_{\mathrm{B}}T}\right)^{\frac{3}{2}} + \dots 
\right) \nn \ .
\eea
In particular, rewriting it in terms of $L$ and using (\ref{e:total_energy_Delta_g})
for $E$ and (\ref{e:N_n_toymodel}) for $\mathrm{N}$ 
\bea
\label{e:E_L}
\!\!\!\!\!\!\!\!\!\!\!\!\!\!\!\!\!\!\!\!\! 
\frac{2}{3}\cdot \frac{\int_M |\nabla \psi|^2dV_g}{\int_M|\psi|^2dV_g} = \frac{1}{L^2}
\left(1 -
\frac{\sqrt{\pi}}{6} \frac{A_h}{V_g(L)} L - \frac{2}{3}\frac{a_1}{V_g(L)} L^2 
- \frac{b_1}{V_g(L)} L^3 + \dots \right) \ .
\eea
This provides a recursive relation to determine the length scale $L$,
that includes stratification and topography terms
and recovers~\footnote{This justifies
the factor $\frac{2}{3}$ in the definition (\ref{e:L_scale_1st}).
Alternatively, it could have been included in 
(\ref{e:Length_scale_Stat}).} $L$ in (\ref{e:L_scale_1st}) as the zeroth-order.

\subsubsection{Equilibrium condition: first derivatives of the potentials.} Mechanical, chemical and 
thermal
equilibria between subsystems $A$ and $B$,
are characterized by the equality of corresponding intensive variables, respectively $P_g$, $\mu$ and $T$ (or $L$), 
namely
\bea
\label{e:equilibrium_conditions}
\left(P_g\right)_A =\left(P_g\right)_B \ \ , \ \ \mu_A = \mu_B \ \ , \ \ T_A = T_B  \ .
\eea
If these conditions are not satisfied, a change/transfer
of effective volume $V_g$, particles $\mathrm{N}$  or ``heat''
happen between subsystems $A$ and $B$.
In other words, $\int_{A,B} 1\, dV_g $, $\int_{A,B} |\psi|^2 dV_g $ and/or 
$\int_{A,B} |\mathbf{\nabla}\psi|_g^2 dV_g$
must change so that $P_g$, $\mu$ and $T$   balance.  
Extrapolating these toy-model results as qualitative indications in the
$L\ll L^{\mathrm{basin}}_{\mathrm{max}}$ regime of the actual QG model, quantities 
$P_g$, $\mu$ and $T$ become functionals of $\psi$ providing information
about the contact between water masses.
This is the approach to diagnostic tools adopted in 
\ref{s:diagnotic_calculation}.

\subsubsection{Stability conditions: second derivatives of the potentials.} 
\label{s:stability}
Thermo-dynamical stability conditions impose sign conditions 
on certain second derivatives of thermodynamical potentials. For instance, 
for the  free energy $F$ it must hold \cite{Callen85}
\bea
\left(\frac{\partial^2 F}{\partial T^2} \right)_{V_g,\mathrm{N}} \leq 0 \ , \
\left(\frac{\partial^2 F}{\partial V_g^2} \right)_{T,\mathrm{N}} \geq 0 \ .
\eea
The violation of such stability constraints signals the conditions
for a phase transition: the system must re-adapt its
internal structure into new configurations reducing the free energy,
possibly incorporating new elements in its physical description.
Extrapolating again such toy-model results to the proper QG model
(caveats regarding non-extensitivity in long-range interacting systems
are particularly relevant here),
this provides another kind of diagnostics with interest in the
assessment of stratification
and topography in the transition from the QG regime to
other dynamical ocean regimes. The condition $L\ll L^{\mathrm{basin}}_{\mathrm{max}}$
in the toy-model suggests the focus on the transition to submesoscale physics (though
a transition to large-scale ocean dynamics could be envisaged 
for a sufficiently large basin size).
For concreteness, let us mention 
the stability condition provided by the positivity of the
heat capacity at constant volume $C_V$ 
\bea
\label{e:C_V}
C_V \equiv \left(\frac{\partial E}{\partial T} \right)_{V_g,\mathrm{N}} \geq 0 \ ,
\eea
providing another diagnostic functional on $\psi$ 
(its expression can be derived from (\ref{e:E_T})).

\subsubsection{Statistical fluctuations.} 
Statistical mechanics provides directly other flow estimators, 
e.g. statistical fluctuations (closely
related to phase transitions). 
The grand-canonical partition function $\Xi=\Xi[T, V_g,\mu]$ in (\ref{e:grandcanonical})
can be expressed as 
\bea
P_gV_g=k_{\mathrm{B}}T\ln\Xi \ .
\eea
The expression for $P_g=P_g(T, V_g,\mu)$ can be obtained from (\ref{e:PV(T)})
and (\ref{e:mu_QG}). The fluctuation in the number of particles
in the grand-canonical ensemble has the form
\bea
\left(\delta \mathrm{N}\right)^2 = \langle \mathrm{N}^2\rangle - \langle \mathrm{N}\rangle^2
= \left(\frac{\partial^2\ln\Xi}{\partial (\beta\mu)^2}\right)_{(T,V_g)} \ ,
\eea
from which we obtain
\bea
\left(\delta \mathrm{N}\right)^2 = 
\left(\frac{2\pi M_o\alpha_o^2}{k_{\mathrm{B}}T}\right)^{-\frac{3}{2}} e^{\frac{\mu}{k_{\mathrm{B}}T}}V_g
= \frac{1}{\left( 4\pi\right)^{\frac{3}{2}}}\left(\frac{V_g}{L^3}\right)e^{\left(\frac{2}{M_o\alpha^2}\right)\mu L^2 }   \ .
\eea 
We note that topography and stratification do not enter in $\left(\delta \mathrm{N}\right)^2$, since
the latter is independent of heat kernel coefficients (dependence on $\alpha_o$
is inherited from the definition of $\mathrm{N}$).  
Fluctuations in the energy in the canonical ensemble
are likewise expressed as
\bea
\left(\delta E\right)^2 = \langle E^2\rangle - \langle E\rangle^2
= \left(\frac{\partial^2\ln Z}{\partial \beta^2}\right)_{(V_g, \mathrm{N})} \ .
\eea
Explicit expressions can be obtained from (\ref{e:partition function_T}).
Quantities $\left(\delta \mathrm{N}\right)^2$ and  $\left(\delta E\right)^2$
can be used as functionals of $\psi$ to diagnose zones of intense 
activity in the QG flow.

\subsubsection{A comment on the relation to full QG statistical mechanics.}
\label{s:Connection_QG_statmech}
The statistical mechanics approach to QG dynamics is the subject 
of an extensive research effort. Comprehensive accounts of the topic
can be found in \cite{BouchetVenaille12,MajWan06}. In particular, the 
RSM theory \cite{Robert90,Miller90,Robert91,RobSom91} incorporates all dynamical invariants 
(\ref{e:Casimir}) in its treatment and
provides a sound theoretical framework where previous approaches 
(in particular for $2$-dimensional Euler flows) are included as 
appropriate limits \cite{BouchetVenaille12}. This theory not only 
explains successfully qualitative properties of geophysical fluids
(in particular with applications in the ocean, cf. \cite{BouchetVenaille12}), 
but also provides a full quantitative approach to the QG dynamics problem,
namely aiming at the description of the most probable final state
by identifying the relevant macrostates and their probability
(note the difference between this quantitative goal and the 
Cauchy problem resolution in PDE approaches).

In contrast with the RSM theory, the statistical mechanics toy-model discussed
here only considers the conservation of energy, ignoring all other dynamical 
invariants.
The elimination of the corresponding long-range interactions entails
key physical differences with the QG model, in particular 
the extensitivity of the energy in (\ref{e:E_T}), not realized
in fluid dynamical models as the QG one. It also impacts 
the possible non-equivalence of statistical ensembles, in particular 
the physical suitability of the microcanonical ensemble in fluid dynamics over 
the canonical ensemble (e.g. \cite{BouCor10}). Our model, explicitly built on
the (fluid non-appropriate) canonical ensemble, ignores this
issue.
For these reasons~\footnote{Another issue concerns the choice
of $\alpha_o$ in (\ref{e:oceandrum_normalization}), due to
the phenomenological absence of a minimal QG length 
(that can be related to $\alpha_o$). However, the thermodynamical description
offers results independent of $\alpha_o$, once expressed in terms of $L$. 
Therefore $\alpha_o$ can be seen as an intermediary technical tool.}, 
the discussed toy-model is not meant to
address the physical dynamics of  QG flows, as e.g. RSM theory does. Rather
than explaining and/or predict, the model attempts to  
provide a set of systematic tools to monitor data of QG flows in a first rough analysis, 
identifying regions where finer and more powerful tools can be applied.

Having said this, the toy-model can provide some interesting insights.
On the one hand, the exponential decay of QG flow interactions 
in scales much larger than the internal Rossby deformation
radius~\cite{VenBou11,CamDauDau14}, makes the toy-model a progressively better physical approximation
at large distances (of interest precisely in the $L\ll L^{\mathrm{basin}}_{\mathrm{max}}$ regime). In particular,
(finite volume) heat-kernel coefficient corrections might 
account for actual physical QG finite volume effects.
On the other hand, the underlying Fock representation may suggest the avenue
to a (physically realistic) statistical field theory, complementary to
the RSM mean-field theory (and possibly relevant for the coupling with fast modes in
\cite{YouBenJel97,XieVan15}).
This can open a path to import spectral geometry expertise in quantum field theory
(e.g. \cite{FurVas11}). 
Finally, formal treatments of the QG model often study
barotropic QG equations. The present geometric discussion recasts 
the full continuous QG model in a form particularly close to the $2$-dimensional
Euler equations, since the (potential) vorticity is also controlled
by an exact Laplacian. This can help to mimic the strategy for 
$2$-dimensional Euler flows, in particular extending 
large deviation results in \cite{BouCor10}.

\section*{References}


\end{document}